\documentclass[final, sort&compress]{article}

\usepackage{xcolor, colortbl}

\usepackage{lineno,hyperref}
\usepackage{textcomp}
\usepackage{xcolor}
\usepackage{algorithm}
\usepackage[noend]{algpseudocode}
\usepackage{placeins}
\usepackage{graphicx}
\usepackage{collectbox}
\graphicspath{{./}}
\DeclareGraphicsExtensions{.png,.jpg}
\usepackage{hyperref}
\hypersetup{
  colorlinks=true,
  linkcolor=blue,
  filecolor=magenta,   
  urlcolor=cyan,
}
\usepackage{caption,subcaption} 
\usepackage{multirow}
\usepackage{mwe} 
\usepackage[noend]{algpseudocode}
\usepackage{babel}[english]
\usepackage{geometry}
\usepackage{pdflscape}
\usepackage{multirow}
\usepackage[export]{adjustbox}
\usepackage{comment}
\usepackage{ulem} 

\definecolor{PreThi}{RGB}{246, 209, 209}
\definecolor{Bio}{RGB}{180, 227, 239}
\definecolor{Unk}{RGB}{209, 246, 224}
\definecolor{Dra}{RGB}{242, 246, 209}
\definecolor{Dig}{RGB}{252, 222, 178}
\definecolor{PumArr}{RGB}{206, 179, 236}

\makeatletter
\def\algbackskip{\hskip-\ALG@thistlm}
\makeatother
\usepackage{geometry}
\begin{document}

\author{ Alicia Beneyto-Rodriguez, Gregorio I. Sainz-Palmero \footnote{Corresponding author: gregorioismael.sainz@uva.es} \\Marta Galende-Hern\'{a}ndez, Mar\'{i}a J. Fuente \\
Department of System Engineering and Automatic Control\\ School of Industrial Engineering, Universidad de Valladolid, Spain \footnote{Emails: {alicia.beneyto,marta.galende,mariajesus.fuente}@uva.es, jmcuencac@aquavall.es}\\
\\
José M. Cuenca\\
Department of  R$\&$D AQUAVALL \\
Camino Viejo de Simancas 229, 47008 Valladolid, Spain\\
}

\date{}

\title{Applying XAI based unsupervised knowledge discovering for Operation modes in  a WWTP. A real case: AQUAVALL WWTP \footnote{This work was supported by the Spanish Government through the Ministerio de Ciencia, Innovaci\'on y Universidades/ Agencia Estatal de Investigaci\'on (MICIU/AEI/10.13039/501100011033) under Grant PID2019-105434RB-C32.}}

\maketitle

\begin{abstract}
Water reuse is a key point when fresh water is a commodity in ever greater demand, but which is also becoming ever more available. Furthermore, the return of clean water to its natural environment  is also  mandatory. Therefore, wastewater treatment plants (WWTPs) are essential in any  policy focused on these serious  challenges.

WWTPs are complex facilities which need to operate at their best to achieve their goals. Nowadays, they are largely monitored, generating large databases of historical data concerning their functioning over time. All this implies a large amount of embedded information which is not usually easy for plant managers to assimilate, correlate and understand; in other words, for them to know the global operation of the plant at any given time. At this point, the intelligent and Machine Learning (ML) approaches can give support for that need, managing all the data and translating them into manageable, interpretable and explainable knowledge about how the  WWTP  plant is operating  at a glance.

Here, an eXplainable Artificial Intelligence (XAI) based methodology is proposed and tested for a real WWTP, in order to extract explainable service knowledge concerning the operation modes of the WWTP managed by AQUAVALL, which is the public service in charge of the integral water cycle in the City Council of Valladolid (Castilla y Le\'on, Spain). By applying well-known approaches of XAI and ML focused on the challenge of WWTP, it has been possible to summarize a large number of historical databases through a few explained operation modes of the plant in a low-dimensional data space, showing the variables and facility units involved in each case.

\end{abstract}

\begin{center}
\end{center}

\section{Introduction} 
\label{sec:introduction}

\begin{figure}[t]
		\centering
		\includegraphics[width=15cm]{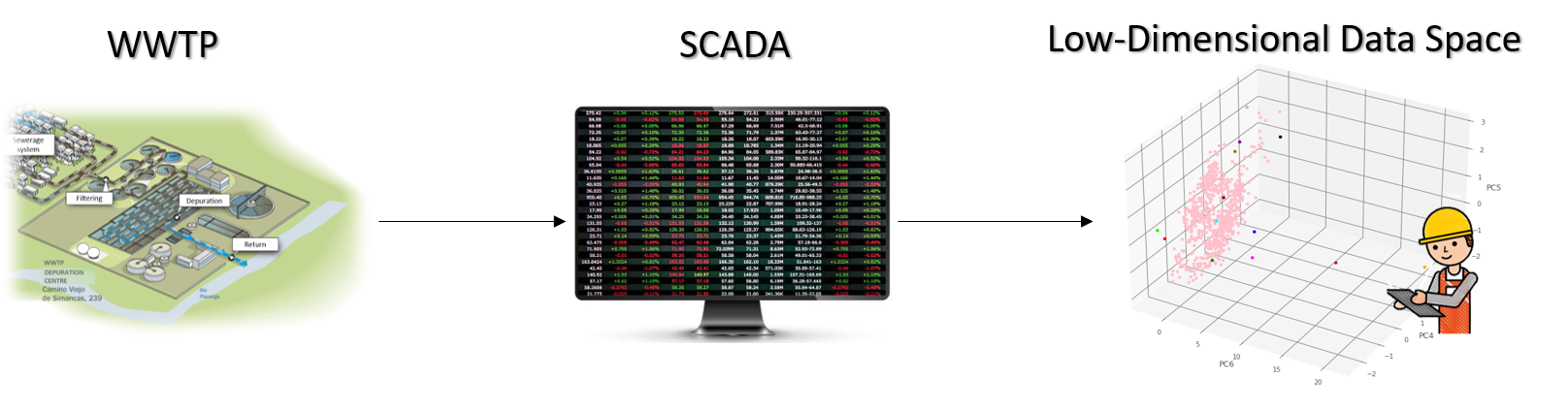}
		\caption{Target overview.} 
		\label{fig:idea}	
\end{figure}

Wastewater treatment plants (WWTPs) are dynamic and complex systems operating  under varying operations, due to fluctuations in influent characteristics, environmental factors and operation adjustments. Understanding the different operation modes within a WWTP is essential for efficiently optimizing the treatment, reducing energy consumption, and ensuring compliance with environmental regulations. However, due to the complexity of the nature of WWTP processes, traditional analytical approaches can often fall short in interpreting these operation  modes. On the other hand, these facilities are monitored in large scale supplying large databases about plant operation, which implies a need for a data summary to simply explain the WWTP operation mode and which can be available at all times for the managers (see Fig. \ref{fig:idea}).

Data-driven techniques have enabled a more systematic analysis of WWTP operations, using real-world plant data collected from monitoring WWTP systems. Some publications address that, taking into account different aims, such as Machine Learning (ML) for predicting wastewater quality in three WWTPs in China \cite{Ly2022}, an evolutionary data-intelligence model, integrating the Extreme Learning Machine (ELM) with Kernel Principal Component Analysis (KPCA) to predict the performance of a WWTP in Nigeria \cite{Abba2020}, or the development of a prediction using the Wastewater Quality Index (WWQI) for a regional WWTP in Malaysia, and Principal Components Analysis (PCA) to handle high-dimensional process data \cite{Rahmat2022}. 

Clustering approaches are incorporated in \cite{Borzooei2019}, combining a $K-means$ based technique with a process simulation model to assess the impact of changing climatic operations on the influent characteristics of a WWTP in Italy. On the other hand, \cite{Han2024} develops an autoencoder based deep clustering method (SMEL) to detect abnormal data with no labels in a WWTP in China. All these approaches in real cases have a lack of explainability; moreover, they do not address the global operation of the plant and neither do they take into account the data from the monitoring systems, which is an essential and mandatory goal for any manager in charge of such facilities.

Here, the target is to know the operation modes of the WWTP facility using its monitoring data collected over time, and then to locate at a glance every new monitored data sample regarding those known, or discovered, operation modes of the WWTP plant. This will also permit the visualization of the large database of the WWTP, containing thousands of variables and records, as a point in the reduced dimension space of the known operation modes; all of this being explained by the plant variables involved in every mode and case. A density based algorithm, DBSCAN (Density-Based Spatial Clustering of Applications with Noise),  is used to discover the operation modes. Before that, a dimensionality reduction is carried out so as to be able to manage the most relevant information in a reduced input space, in order to deal with the \textsl{curse of dimensionality} for the ML techniques and to improve the explainability and interpretability of the outcome. Therefore, expert knowledge, principal component analysis (PCA), and Shapley Additive Explanations (SHAP) are used. 

Some works have shown the effectiveness of DBSCAN in clustering for environmental and industrial process data, such as \cite{Solano2022} to localize illegal industrial discharges of polluting wastewater in sewer networks. However, as DBSCAN effectively groups similar operation states, it does not inherently provide any explainibility about the generated clusters in the domain terms. To bridge this gap, SHAP is applied to explain the contribution of individual features to the identified operation modes. This technique is the most popular within the scope of the  eXplainable Artifical Intelligence (XAI) methods \cite{Cohen2024} \cite{Clement2024} \cite{Dwivedi2023} \cite{Cilinio2023}. The concept of explainability is becoming ever more relevant, even mandatory when AI methodologies are applied to different fields such as \cite{EUActa} \cite{Panigutti2023} \cite{Eth2019}, but water systems are also considered a critical domain \cite{Polaine2022} \cite{Gamache2021}.

According to the goals for this work described above, the main contributions of this work are as follows:

\begin{enumerate}
	\item An explainable  machine-learning based methodology is created to summarize large and high dimensional WWTP monitoring databases as a point in a reduced space of operation modes.
	\item This approach allows the current WWTP operation mode regarding the known modes to be located at all times.
	\item Every WWTP operation mode  is explained by the variables of the WWTP plant.
	\item The validation of clustering results through expert knowledge, ensuring that the detected operation modes are meaningful and useful for WWTP operators in practice.

\end{enumerate}

The rest of the paper is organized as follows. In Section \ref{sec:introduction}, the introduction provides the context and motivation for the study, highlighting the challenges in monitoring WWTPs and the importance of developing explainable, data-driven methods for operation mode detection, as well as a brief state of art in this area. Section \ref{sec:foundations} provides the theoretical foundations necessary for this work, offering a brief overview of PCA for dimensionality reduction, DBSCAN for clustering, and SHAP for model interpretability. Next, Section \ref{sec:proposal} describes the proposed methodology, while in Section \ref{sec:casestudy}, a real case study is presented, where the methodology is applied and the results are analyzed, validating the clustering results through expert knowledge. Finally, Section \ref{sec:conclusions} summarizes the main findings, discusses the implications of the study, and outlines potential directions for future research.

\section{Some Foundations}
\label{sec:foundations}

The following subsections briefly introduce the theoretical items supporting this work.

\subsection{Principal Component Analysis (PCA)}
\label{sec:pca}

AQUAVALL data records  contain a large number of variables with a high probability of correlation. This large number of variables (data dimensionality) is a challenge both for their computation  and to know the specific relevance of each one. In this work, Principal Component Analysis (PCA) \cite{Cadima2016} \cite{Jollife2002} is used to discover the relevance of each one, making a dimensionality reduction under some conditions, thus transforming the original variables into a new set of extracted variables which can be ranked by the data variance of each one. Then a  selection can be  carried out to create a subset  under some operations, while preserving most of the information with a small number of variables (principal components).

The goal behind the PCA is to be able to achieve a low-dimensional representation of the data, preserving as much information as possible. This is achieved by identifying the directions (principal components) along which the data vary the most. The first component accounts for the largest possible variance, with each subsequent component capturing the highest remaining variance, while being orthogonal to the preceding components. 

Mathematically, PCA \cite{Reddy2020} begins by standardizing the dataset to ensure that each variable contributes fairly  to the analysis. Given a dataset $X$ with $n$ samples and $m$ variables, standardization is performed by subtracting the mean and dividing by the standard deviation:
\begin{equation}
X = \frac{X_{raw} - \mu}{\sigma}
\end{equation}
where $\mu$ is the mean vector and $\sigma$ is the standard deviation vector.

Next, the covariance matrix $C$ is computed as:
\begin{equation}
C = \frac{1}{n-1} X^T X
\end{equation}
This matrix captures the relationships between variables. The principal components are then determined by solving the eigenvalue problem:
\begin{equation}
C v = \lambda v
\end{equation}
where $\lambda$ represents the eigenvalues and $v$ the corresponding eigenvectors. The eigenvectors (principal components) are ranked  based on descending eigenvalues in order to be able to select the most significant components.

The raw data are transformed into an n-dimensional space, permitting a k-dimension subspace to be selected by choosing the k top eigenvectors of the covariance matrix, and therefore the k top principal components. These eigenvectors form the new basis for the data. The full transformation for all data points can be written as:
\begin{equation}
X_{new} = X V_k
\end{equation}
where $X_{new}$ is the transformed dataset in the reduced-dimensional data space, and $V_k$ consists of the top $k$ eigenvectors.

However, this approach has a serious disadvantage: a lack of interpretability concerning their principal components and their meaning with regard to the problem domain.

\subsection{DBSCAN}

DBSCAN (Density-Based Spatial Clustering of Applications with Noise)  \cite{Deng2020} \cite{ester1996}  defines clusters as dense regions of points separated by areas of lower density. It does this by classifying points as core points, border points, or noise, based on two main parameters: $\varepsilon$ $(epsilon)$, which determines the neighborhood radius, and $MinPoints$, the minimum number of points required to form a dense region. This density-based approach makes DBSCAN highly effective for applications ranging from anomaly detection to spatial data analysis. Moreover, its flexibility, robustness to noise, and ability to discover non-spherical clusters make it a widely used and valuable clustering algorithm in many real-world applications.

The $\varepsilon$-neighborhood of a point $p$ is defined as:
\begin{equation}
N_{\varepsilon}(p) = \{ q \in D \mid d(p, q) \leq \varepsilon \}
\end{equation}
where $D$ is the datset, $d(p, q)$ is the distance function (typically Euclidean distance), and $\varepsilon$ is the neighborhood radius.

A point $p$ is considered a core point (points that have at least $MinPoints$ other points within a distance of $\varepsilon$) if it satisfies the density operation:
\begin{equation}
    |N_{\varepsilon}(p)|\geq \mathit{MinPoints}
\end{equation}
where $|N_{\varepsilon}(p)|$  represents the number of points in the $\varepsilon$-neighbourhood of $p$.

On the other hand, a point $p$  is classified as noise if it is not density-reachable from any core point, meaning it does not belong to any cluster:
\begin{equation}
    |N_{\varepsilon}(p)| < \mathit{MinPoints}
\end{equation}

Therefore, DBSCAN groups all density-connected points into clusters, while points that do not meet the density criteria are labelled as noise points. This term, noise, is not considered for this work as we are searching for events which could have happened once time and they can not be considered as noise.

\subsubsection{DBSCAN Tuning}

Clustering algorithms, unlike supervised learning,needs some methodology for tuning its own parameters in order to achieve success. Unlike traditional clustering methods, that assume clusters have a specific shape or require a predefined number of clusters, DBSCAN detects groups of varying structures and effectively handles noise within a dataset. Its effectiveness depends on the  two  parameters commented in the previous section: $MinPoints$ and $\varepsilon$ $(epsilon)$. There is no definitive formula to determine the ideal $MinPoints$ or $\varepsilon$  values; however, its selection should be guided by the nature of the dataset and expert knowledge. 

On the other hand, the search for a fair $\varepsilon$ value also has to be addressed. One effective and well-known way to estimate $\varepsilon$  is by analyzing the distances between each point and its $k$-nearest neighbors \cite{Sander1998}, where $k$ corresponds to the chosen $MinPoints$. By checking out  these distances in ascending order on a $k$-distance graph, the optimal $\varepsilon$ can be identified at the point of highest curvature, often referred to as the \textsl{''elbow point''}. In this way, it is possible to distinguish cluster boundaries by selecting a threshold that separates dense regions from sparse areas.

\subsection{Shapley Additive Explanations (SHAP)}
\label{sec:shap}

Nowadays, eXplainable Artificial Intelligence (XAI) is a goal, if not a mandatory goal, for AI applications. One of the options for the said target is the Shapley Additive Explanations (SHAP) method \cite{Lundberg2017}. This is a powerful XAI method for explaining the output of machine learning models.  It provides a way to attribute the prediction of a machine learning model to its individual input features in a fair and consistent manner, providing an alternative for addressing  the classic and well-known  black box challenge. SHAP values  are an adaptation  from the Shapley values, used in cooperative game theory, for explaining machine learning models.

Shapley values \cite{Hart1989} were originally introduced in the context of cooperative game theory, where the goal was to fairly distribute the 'payout' (or value) generated by a team of players. In the case of machine learning, the 'game' involves predicting an outcome, and the 'players' are the individual features, or variables,  of the model. Shapley values provide a way of assigning a contribution to each variable based on its impact on the prediction. The Shapley value for each variable in a machine learning model is calculated by evaluating its contribution to all possible combinations of variables. Given a predictive model $f$ and a set of $m$ variables, the Shapley value for a $variable_i$ is calculated as:
\begin{equation}
\phi_i = \sum_{S \subseteq N \setminus \{i\}} \frac{|S|!(m - |S| - 1)!}{m!} \left[ f(S \cup \{i\}) - f(S) \right]
\end{equation}
where $S$ is a subset of variables excluding the variable $i$, $f(S)$ is the model's prediction when using only the features in $S$, and $f(S \cup $i$)$ is the model's prediction when feature $i$ is added.

However, SHAP simplifies this process by providing an efficient and scalable way to calculate Shapley values for machine learning models. Several versions of SHAP were introduced by \cite{Lundberg2017}: DeepSHAP, KernelSHAP, LinearSHAP and TreeSHAP, developed for different types of models. In this work, KernelSHAP has been used to explain PCA and clustering outputs, as it is a model-agnostic method for estimating SHAP values efficiently, approximating Shapley values using a weighted linear regression approach, has been used.

\section{Methodology}
\label{sec:proposal}

\begin{figure}[t]
		\centering
		\includegraphics[width=15cm]{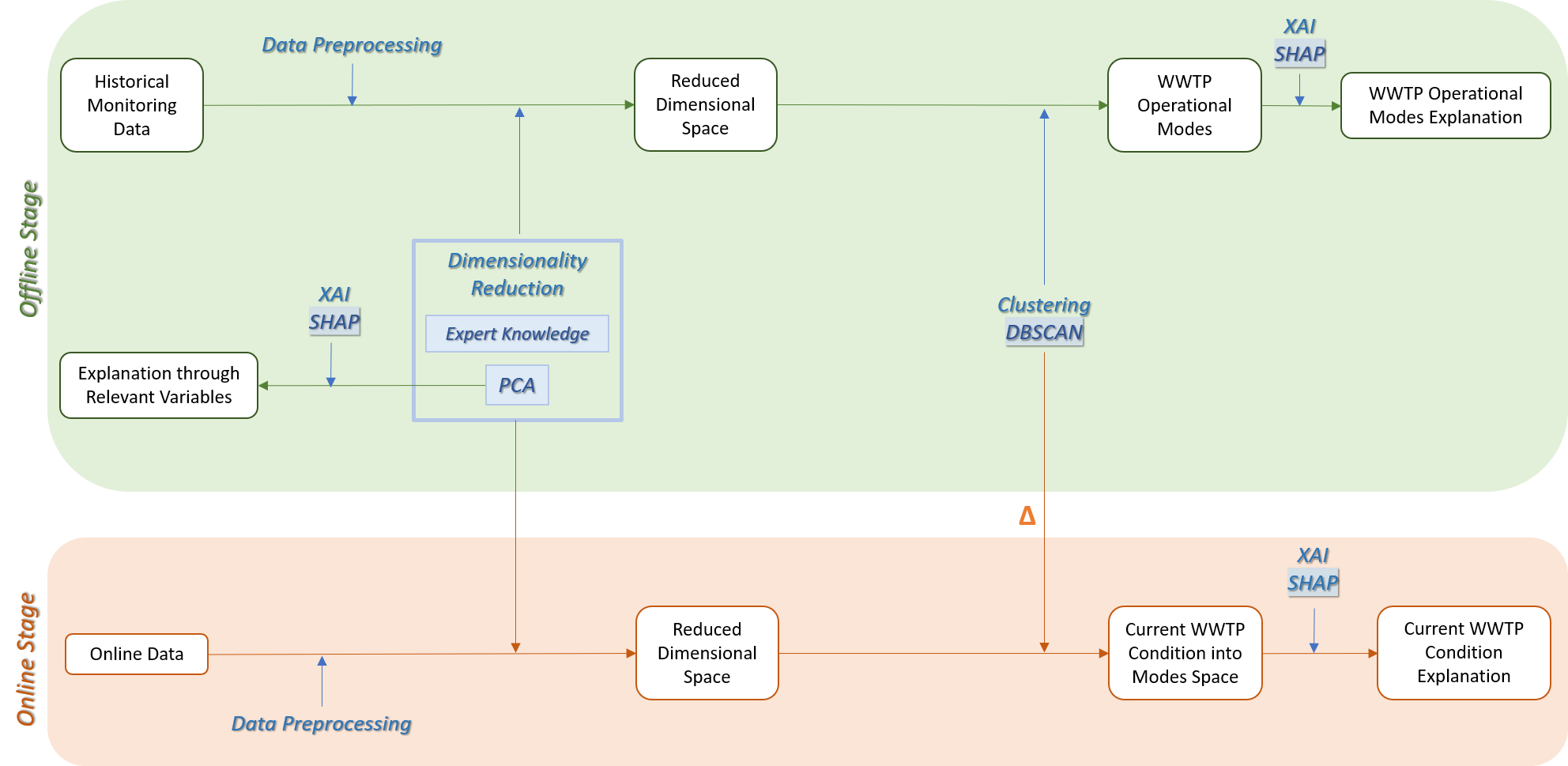}
		\caption{Methodology scheme.} 
		\label{fig:methodology}	
\end{figure}

The main goal of this proposal is to understand at a glance the operation mode of the WWTP plant  regarding its known standard operation modes at any time, and all this is provided in a reduced space of variables in an  explainable way for the WWTP managers and operators. Consequently, this implies summarizing the large amount of high-dimensional data provided by the WWTP monitoring systems, explaining the main roots (variables) involved for each standard and non-standard operation modes. All this is based on a data-driven solution, so a sufficient quantity and quality of data, is mandatory for that goal. The proposed methodology (see Fig. \ref{fig:methodology}) is based on a dimensionality reduction through expert knowledge and  PCA, clustering by DBSCAN and explanations by SHAP, involving two main, usual stages for its setup: offline and online. All this is detailed in Sec.  \ref{secmethoexp}

\subsection{Offline Stage}

A first and critical methodological step is focused on the preprocessing  of the historical monitoring data to be used: data cleaning, resampling and standardization. 

The modern facilities of monitoring and control for this type of plants capture thousands of variables of different nature and different goals using a SCADA system which  can contain several types of error such as  missing values, outlayers, etc., so a cleaning procedure is mandatory. Furthermore, the number of variables is usually too large, containing \textsl{useless} variables, such as those regarding the computer platform or similar  ones. So, this challenge has to be addressed and solved, removing SCADA variables used for other purposes, while leaving the variables monitoring the WWTP process that are of interest for our goals, always based on the expert-knowledge of the plant. Then, another critical point is the sampling time, which  has consequences over the outcome quality and its reliability/usability, but  also over the computational efforts, so a resampling step must be addressed. Finally, the last part of the preprocessing involves the standardization of each of the acquired variables. Here, the \textsl{z-score} formulation is used for this goal.

Notwithstanding the above, the most relevant challenge to address when a data driven approach is involved is the data dimensionality, which is usually done through variable selection or extraction to manage the \textsl{curse of dimensionality} \cite{Anovar2021}. Then an extraction of variables/features is carried out to obtain new variables, concentrating the information about the WWTP. Here, the well-known Principal Components Analysis (PCA) \cite{Greenacre2022} is involved. Nevertheless, another challenge appears as a consequence of this technique: the interpretability and explainability regarding the original variables is lost. This issue must be addressed, so SHAP is used for this goal. The dimensionality reduction from the principal components is based on two main criteria: the variability preserved in the remaining new extracted variables  and the interpretability of the outcomes to be managed by the technicians in charge of the WWTP plant.

When the new input space is defined, then the stage devoted to finding operation modes is  launched. Here, a well-known clustering approach is used, based mainly on density: DBSCAN.
This algorithm is tuned in order to obtain the best data partitioning possible, the clusters, or in this case the WWTP operation modes, being explained  by SHAP values in terms of WWTP variables. This XAI technique shows the contribution, or relevance, of each variable for each cluster, and thus for each operation mode of the WWTP plant.

\subsection{Online Stage}

This stage involves putting in place the outcome from the previous stage. In modern plants, to know the current operation of the plant at any particular moment is a mandatory task for managers. Taking the information available from the monitoring system is the most usual option, showing the values on a screen, as with a standard SCADA application; here, however, these data are turned into a point in the low-dimensional data  space of the WWTP operation modes, showing the current operation of the WWTP plant regarding the operation modes known so far, as well as having knowledge of the relevant variables of that mode. 

The first step is based on the preprocessing of the new monitoring data, in the same way as in the offline stage: data cleaning, resampling and standardization. Then, after the dimensionality reduction based on the principal components obtained from the previous stage, an incremental clustering (Incremental DBSCAN) must be carried out, incorporating the new data for updating the operation modes of the plant and to compare the evolution of the plant regarding the previous operation modes. Finally, these modes are explained by SHAP.

If the original historical data were considered outdated due to the plant's evolution or after a long period of time, the offline stage would be restarted for a better performance.

\section{Case Study: AQUAVALL WWTP}
\label{sec:casestudy}

\begin{figure}[h]
		\centering
		\includegraphics[width=15cm]{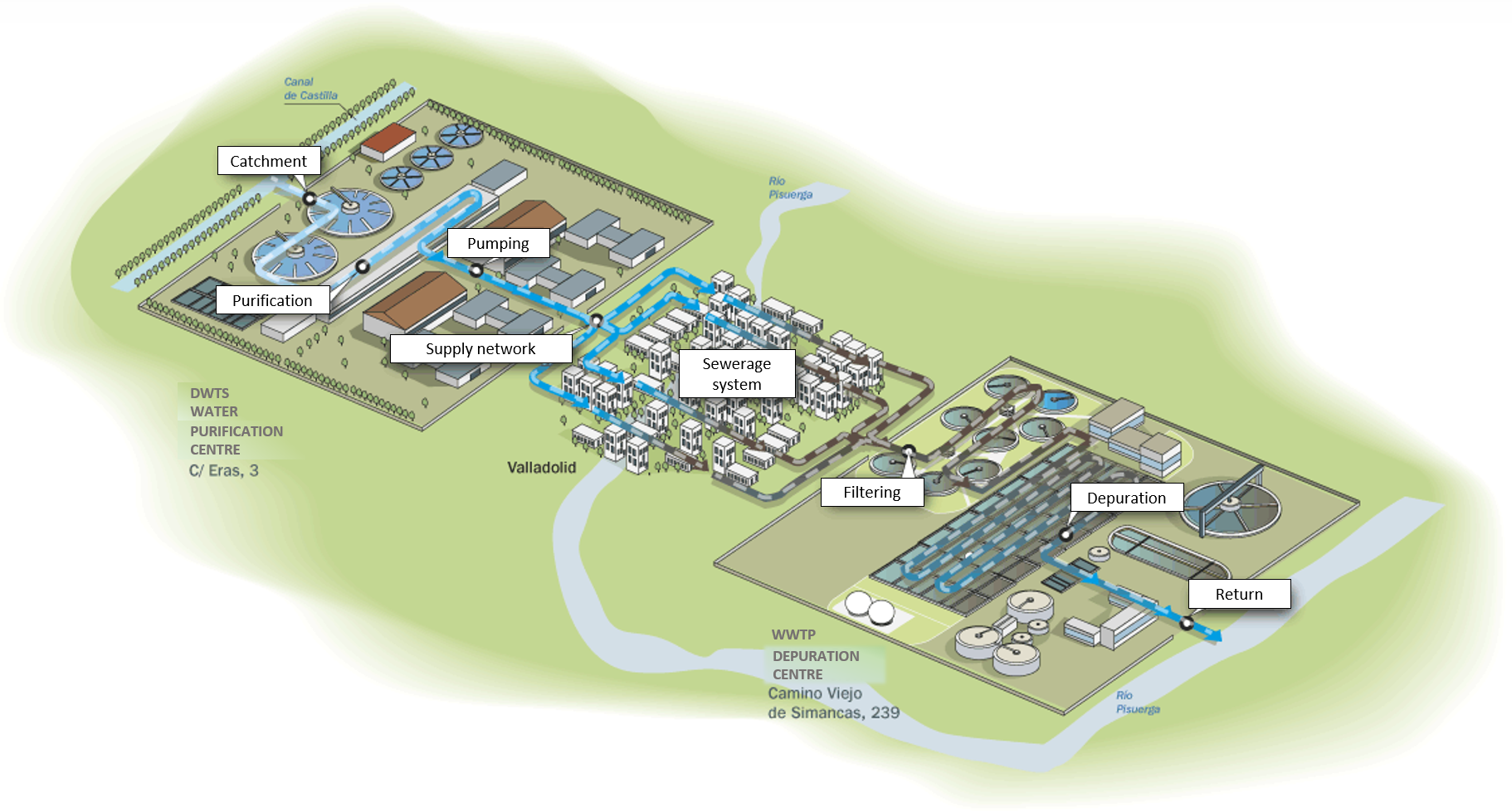}
		\caption{  Water cycle at  AQUAVALL - Valladolid (www.aquavall.es).} 
		\label{fig:ciclo}	
\end{figure}

\begin{figure}[h]
	\begin{center}
		\includegraphics[height=5cm]{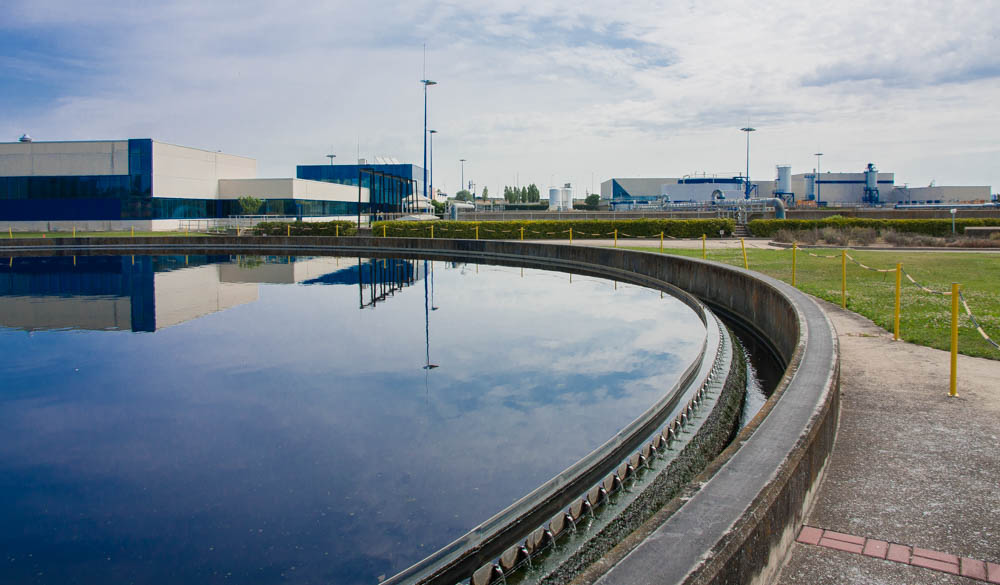}
		\caption{AQUAVALL WWTP - Valladolid (www.aquavall.es).} 
		\label{fig:edar}

	\end{center}
\end{figure}

The real WWTP used for this work as case study is run by AQUAVALL \footnote{https://aquavall.es}. AQUAVALL is the public consortium in charge of the integrated water system in the city of Valladolid (Castilla y León, Spain) (see Fig. \ref{fig:ciclo}). This  consortium is in charge of the integrated water service for more than 344,600 households, industries, and other water consumers under optimal quality operations. The main facilities for these goals are two drinking water treatment plants (DWTP), a wastewater treatment plant (WWTP), and a laboratory located within one of the DWTP plants, where both the quality of drinking water and wastewater discharges into the network are monitored. Water is delivered through a network of pipelines spanning 639 $km$. Meanwhile, the wastewater generated in the city area  reaches the WWTP via a 786 $km$ sewer system. This WWTP processes around 122,000 $m^3$ of wastewater daily, being able to manage a maximum flow rate up to $\frac{m^3}{seg}$, equivalent to the organic load generated by 570,000 people. 
The AQUAVALL WWTP (see Fig. \ref{fig:edar}) is organized into three processing lines: water, sludge, and gas, which are arranged to keep interconnected processes in close proximity. The water treatment line consists of several stages: an intake well, a storm-water tank, preliminary treatment, primary sedimentation, and a biological treatment process. A brief overview of all this is available at \cite{AquavallCapta2025}.

\subsection{Experimental Methodology}
\label{secmethoexp}

The proposal described in Sec. \ref{sec:proposal} has been implemented using real raw data  obtained from the  monitoring system of AQUAVALL WWTP, embracing a time period of 15 months with a sampling time of 1 minute (658,084 data samples). Each sample records 3,231 variables from the systems involved in the different sides of the WWTP facilities. Twelve months of data were used for the offline stage, and 3 months of data for the online stage.

\subsubsection{Offline Stage}

\vspace{0.5\baselineskip}
\textbf{$\bullet$ Data preprocessing}
\vspace{0.5\baselineskip}

\label{sec:preproces}
This huge number of variables also implies a compulsory procedure of cleaning and removing of variables which do not concern the wastewater treatment, such as  security systems and others. There are also variables with value $0$ over the entire time, so they were removed, as well as variables with a level of  missing values higher than $1\%$. In the opposite case related to missing values, the average between samples, the previous value or the next value have been applied to deal with the issue. Other variables concerned the cumulative operation hours of devices deployed throughout the whole facility, such as electrical engines, sensors, valves, etc. All these variables were also removed because of their irrelevance for the WWTP operation. Finally, the number of remaining variables to be managed for the proposed goal were 453. This is a first step in the data dimensionality reduction.

On the other hand, the raw data were sampled every minute, which  implies some issues for dealing with the goal of this work, in both a very high computational load and sampling time, one minute being a very low for this WWTP process and its management. So, a resampling of data is needed  to address both  issues. In this proposal, two resampling times have been considered based on domain expertise:
\begin{itemize}
    \item \textsl{12 hours}: considering the 12 hours for resampling as 08:00-20:00 and 20:00-08:00. 
    \item \textsl{24 hours}: considering the 24 hours for resampling as 00:00-24:00. 
\end{itemize} 

A standardization is then carried out using the well-known standard score, or $z-score$, for the variables at each new resampling time.

\vspace{0.5\baselineskip}
\textbf{$\bullet$  Data dimensionality reduction}
\vspace{0.5\baselineskip}

As for the processing of the raw data, some of the available variables were removed due to different issues, remaining 453 variables. However, this reduction in the number of variables, or data dimension, is not still enough for machine learning approaches and others,  which have to deal with the \textsl{curse of dimensionality}, so a further reduction is compulsory. This is addressed here by the well-known PCA \cite{Cadima2016} \cite{Jollife2002}. In Sec. \ref{sec:pca}, a brief introduction to PCA has been done, showing that this extraction of variables is useful for dealing with the issue of dimensionality reduction. However, two new issues must be addressed:
\begin{enumerate}
			\item  The dimensionality of the new input space, or number of principal components to be kept: here, this number has been managed taking into account the variance accumulated by these components and the complexity of the model. 
			\item The explainability, the techniques for dimensionality reduction often have a serious issue, which is the loss of interpretation and explainability  based on the original variables. Here, this challenge has been addressed by another well-known approach: SHAP \cite{Marcilio2020} \cite{Lundberg2020}.
\end{enumerate}

All this concerns both resampling times.

\vspace{0.5\baselineskip}
\textbf{$\bullet$ WWTP Operation modes: DBSCAN based clustering}
\vspace{0.5\baselineskip}

In this work, we use $MinPoints=1$, meaning that every point is considered a potential cluster core, and clusters can form with a single point. While this setting allows greater flexibility in detecting small or sparse clusters, it may also lead to increased sensitivity to noise.

Once $MinPoints$ han been set, an appropriate $\varepsilon$ value is selected for each sample time and each stage of the methodology, using the $k$-nearest neighbors proposal mentioned in Sec. \ref{sec:proposal}.

\vspace{0.5\baselineskip}
\textbf{$\bullet$  Explanation of the operation modes: SHAP based XAI}
\vspace{0.5\baselineskip}

Every cluster, or operation mode, discovered in the previous stage is explained by the most contributive WWTP variables through the SHAP values, showing these contributive variables and their corresponding WWTP units.

\vspace{0.5\baselineskip}
\textbf{$\bullet$ WWTP operation modes into a reduced input space}
\vspace{0.5\baselineskip}

The location of the WWTP operation modes in the input reduced space, according to the most relevant variables,  shows the plant's known modes, and their closeness or similarity, to the managers and technicians in charge of its operation, as well as its explanation through its more contributive variables of the WWTP and the plant units involved.
 
\subsubsection{Online Stage}
\vspace{0.5\baselineskip}
\textbf{$\bullet$ WWTP operation mode at $t_i$}
\vspace{0.5\baselineskip}

At every sampling time, new data are supplied by the monitoring system and, after its processing according to the previous steps, its prediction regarding to the previously known modes is made and visualized, explaining the more contributive variables for this current operation mode. Here, the plant manager sees the current operation of the WWTP plant and its closeness to the known operation modes of the plant happening over time.   

\vspace{0.5\baselineskip}
\textbf{$\bullet$ Model updating/rebuilding}
\vspace{0.5\baselineskip}

The model generated must be updated, or rebuilt, by incremental clustering based on DBSCAN  to see the evolution of the plant over time.  Here, the last 3 months of data are used for this goal, following the previous stages when necessary. On the other hand, when the WWTP mode model is considered updated, all of the methodology can be carried out for a new mode model.

\subsection{Results and Analysis}
\label{sec:resultados}

The experimentation carried out in this work is based on 15 months of WWTP data: 12 months (from 2022/08/01  to 2023/07/31) are used for  obtaining the mode model  by tuning parameters,  and 3 months (from 2023/08/01 to 2023/10/31) for simulating the online stage, considering in total 658,084 data samples from 3,231 monitored variables, with 525,600 and 132,484  samples for building and testing, respectively.

These data were processed in order to remove useless variables and other issues, such as those with constant values, without impact in our goals, or with a relevant number of missing values, as mentioned previously. Here, 2,393 variables suffered a rate of missing values lower than $1\%$,  with an average rate of $0.25\%$, and 838 variables had a higher missing value rate than $1\%$, showing an average of $41.06\%$.  Finally, the number of remaining variables of interest for this work  were 453. 

The following step was data resampling. Data were originally sampled at $1 min$; however, this sampling rate generates some difficulties that need to be fixed: such an amount of data implies large computing resources and high consuming time, but this sampling time is also not very suitable for the WWTP dynamics or the plant operation expected by managers. Two data resamplings were carried out according to the available expert knowledge: 12 and 24 hours, as described in Subsec. \ref{sec:preproces}. The results of these resamplings are shown in Table \ref{tab:resampling}.

\begin{table}
\begin{center}
	\begin{tabular}{||c|c||c|c||} \hline
	Offline 	&Sampling Time 	& Samples 	& Variables  \\ \hline 
			& 24 hours		& 365		&453 		\\ \cline{2-4}
			& 12 hours		&730 		&453			\\ \hline \hline
	Online 	&Sampling Time 	& Samples	& Variables \\ \hline 
			& 24 hours		& 91		&453 		\\ \cline{2-4}
			& 12 hours		&183 		&453			\\ \hline \hline
	
	\end{tabular}
\end{center}
	\caption{Data resampling: $T_s=12 hours$ and  $T_s=24 hours$.}
	\label{tab:resampling}
\end{table}

This reduction in the number of variables, or dimensional input space, is highly significant, but is not enough to be applied successfully to machine learning techniques. So a process of  dimensionality  reduction is mandatory. This reduction in dimensionality was carried out by PCA, a well known approach addressing the issue of the number of principal components to be kept in each case.

\begin{figure}[ht]
		\centering
    		\makebox[\textwidth][c]{
		\begin{tabular}{cc}

				\includegraphics[height=4cm]{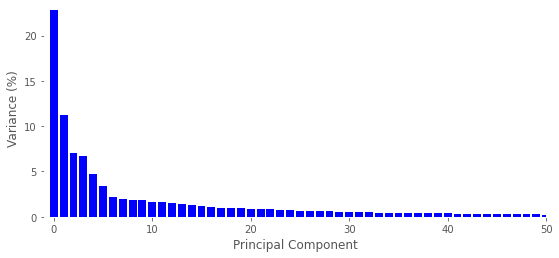} & \includegraphics[height=4cm]{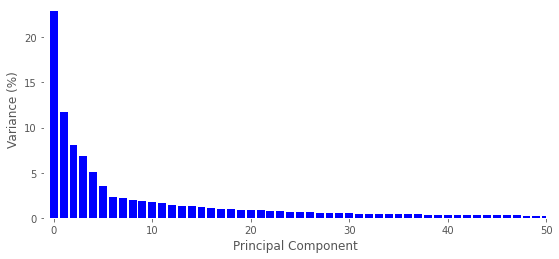}\\
				(a) Principal Component Variance $T_s=12 hours$	\hspace{0cm}			&(b) Principal Component Variance $T_s=24 hours$		
		\end{tabular}
 		 }
	\caption{Relevance of the Principal Components by PCA.}	
	\label{fig:pca}
\end{figure}

Fig. \ref{fig:pca} (a) and (b) show the variance  supplied by each principal component for $T_s= 12hours$ and $T_s=24 hours$, respectively. Here,  the selection of the $m$ principal components was based on the 'elbow' of the graphics, taking into account the variances preserved by this component selection and its number. In both cases, the number of principal components was 7, in the case of $T_s=12hours$, preserving $58,04\%$ of variance, and for $T_s=24 hours$, $60,51\%$. Therefore, this data dimensionality is reasonable for dealing with matching learning approaches.

However, when using these new extracted variables, or principal components, there is a lack of explainability; so we use SHAP to know the most relevant WWTP variables (i.e., the first ten) for the selection of the main principal components, as displayed in Table \ref{tab:SHAPPCAOffline}, and the relevant  WWTP facilities/units (see Table \ref{tab:SHAPPCAInstalaOffline}).

\begin{table}[h]
    \centering
    \begin{tabular}{||c||c||}\hline
       \multicolumn{2}{||c||}{WWTP Relevant Variables} \\ \hline
        \textbf{$T_s=12 Hours$} 						 		& \textbf{$T_s=24 Hours$} \\ 
        \hline \hline
        \cellcolor{PreThi}BP3042\_V.K\_Factor 					 & \cellcolor{Bio}RE2081\_SA\_SAF.val  \\   \hline
       \cellcolor{PreThi} BP3041\_V.K\_Factor  					 &\cellcolor{PreThi} BP3042\_V.K\_Factor \\  \hline
        \cellcolor{Bio}FE201.VSca  							&\cellcolor{Bio} FE201.VSca \\  \hline
        \cellcolor{Unk}ZHRecirculacionFangosBiologico3.TotalActual 		 & \cellcolor{Unk}ZHFangosExceso.TotalActual \\  \hline
        \cellcolor{Dra}Escurridos.PT100.VALOR  					& \cellcolor{PreThi}BP3041\_V.K\_Factor \\  \hline
        \cellcolor{Bio}FR204\_EA\_IA.VSca  					&\cellcolor{Dig} IQ4103.VSca \\  \hline
        \cellcolor{Bio}RE2081\_SA\_SAF.val  					& \cellcolor{Bio}IP223\_EA\_IA.VSca \\  \hline
        \cellcolor{Bio}FR202\_EA\_IA.VSca  					&\cellcolor{Bio} FR202\_EA\_IA.VSca \\  \hline
        \cellcolor{Unk}ZHEntradaBiologico1.TotalActual  			& \cellcolor{Bio}FR204\_EA\_IA.VSca \\  \hline
        \hline
    \end{tabular}
    \caption{Relevance by SHAP for 7 first Principal Components.}
    \label{tab:SHAPPCAOffline}
\end{table}

\begin{table}[H]
    \centering
    \begin{tabular}{|c l|}
        \hline \hline
	& WWTP Units \\ \hline \hline
        \cellcolor{PreThi} & Preprocessing and Thickening \\  \hline
        \cellcolor{Bio} & Biological \\  \hline
        \cellcolor{Dra} & Drained \\  \hline
        \cellcolor{Dig} & Digestion \\ \hline
        \cellcolor{PumArr} & Pumping Arrival \\ \hline
        \cellcolor{Unk} & Others \\  \hline
        \hline
    \end{tabular}
    \caption{WWTP units involved.}
    \label{tab:SHAPPCAInstalaOffline}
\end{table}

The most influential variables obtained for each resampling time are ranked in a slightly different way, so the sample time influences the results: DBSCAN operation modes as well as SHAP explanations. After the explanation of the 7 principal components, the following stage involves the detection of WWTP operation modes  for each resampling time ($T_s$).

\subsubsection{WWTP Operation Modes}

The detection of the WWTP operation modes is based onthe DBSCAN algorithm, tuned according to the experimental methodology described in Sec. \ref{sec:proposal}. In order to avoid a very long section, only the most relevant DBSCAN results are shown.

\vspace{0.5\baselineskip}
\textbf{ $\bullet$ Case: $T_s=12 hours$}, Offline
\vspace{0.5\baselineskip}

\begin{figure}[h]
	\centering
    	\makebox[\textwidth][c]{
	\begin{tabular}{cc}
			\includegraphics[height=8.5cm]{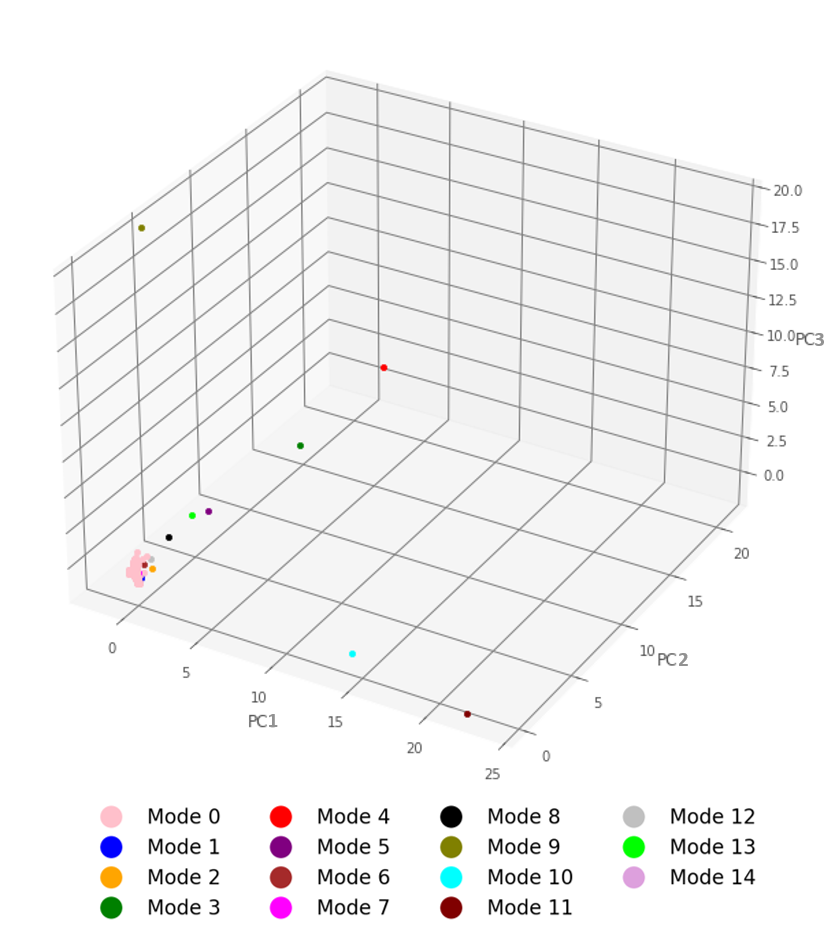} & \includegraphics[height=8.5cm]{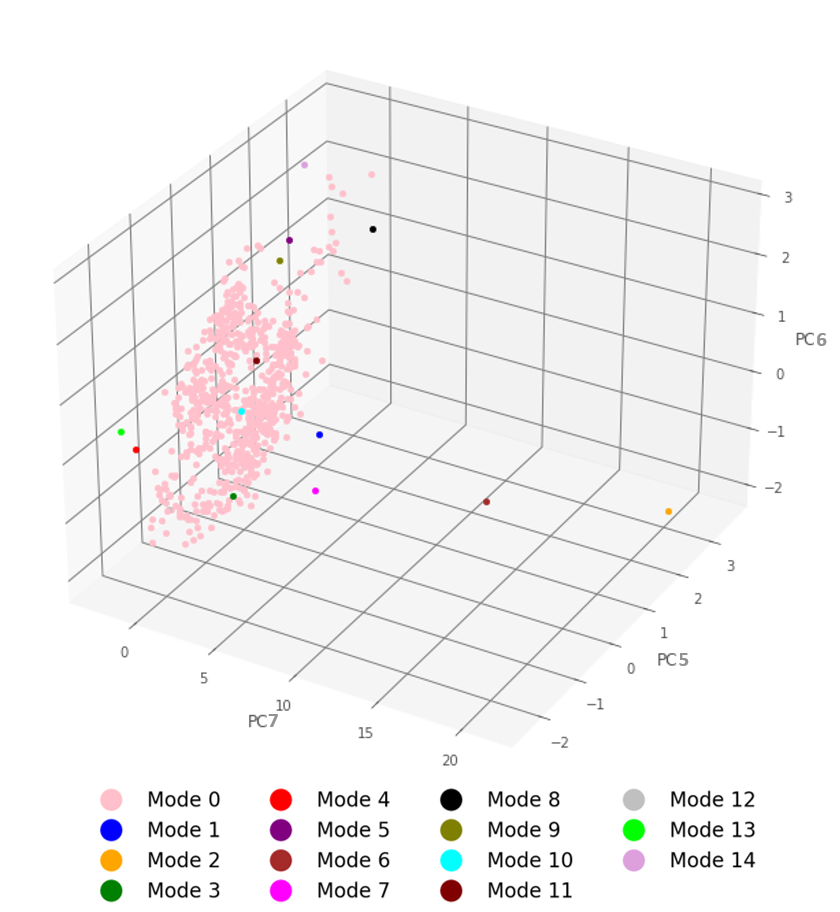}\\
			(a) Operation Modes: view by  $PC1$,$PC2$ and $PC3$.					& (b) Operation Modes: view by  $PC5$, $PC6$ and $PC7$.
	\end{tabular}
	}
	\caption{Operation Modes based on DBSCAN ($\varepsilon=1.3$  and the $MinPoints=1$).}
	\label{fig:dbscan12h}
\end{figure}

\begin{figure}[H]
	\centering
    	\makebox[\textwidth][c]{
	\begin{tabular}{cc}
			\includegraphics[height=8.5cm]{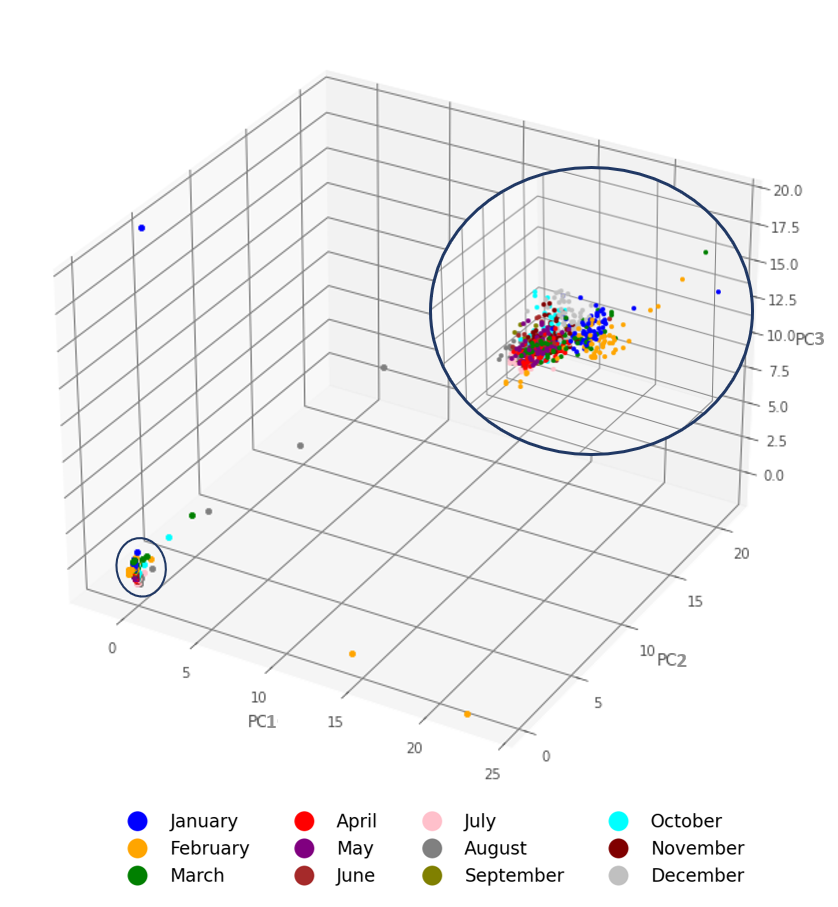} & \includegraphics[height=8.5cm]{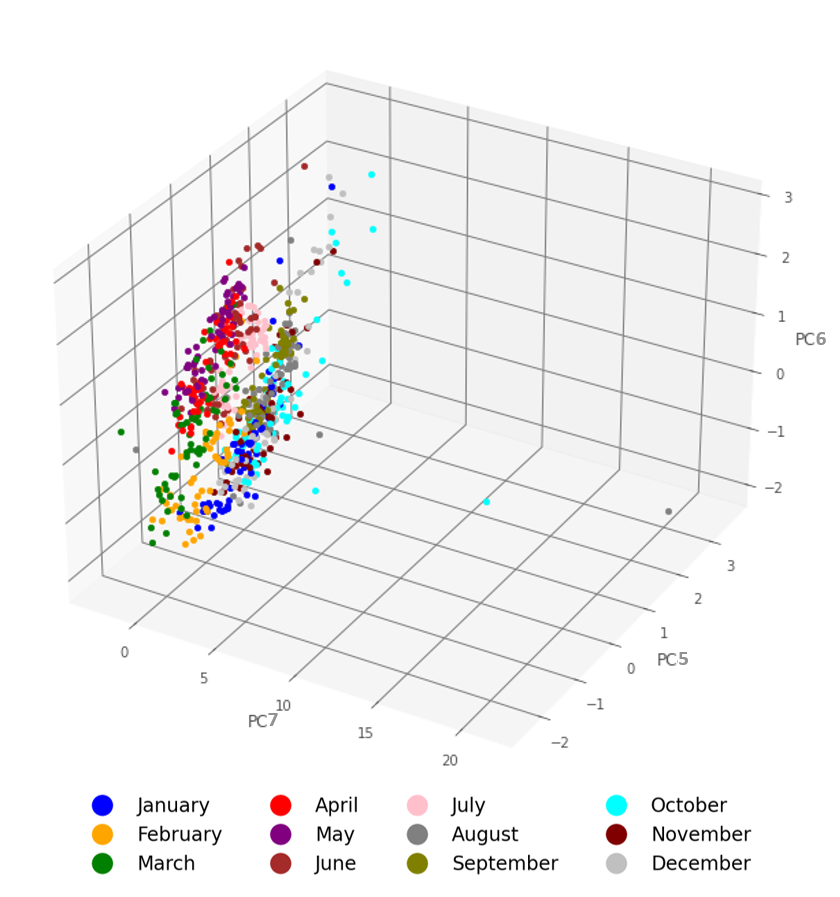}\\
			(a)  Operation Modes by months ($PC1$,$PC2$ and $PC3$).					& (b) Operation  Modes by months ($PC5$, $PC6$ and $PC7$).
	\end{tabular}
	}
	\caption{Operation Modes based on months ($\varepsilon=1.3$  and the $MinPoints=1$).}
	\label{fig:dbscan12hMes}

\end{figure}

Here, the best tuning for the DBSCAN parameters was:  $\varepsilon=1.3$  and the $MinPoints=1$. This last parameter was a hard restriction in order to detect operation modes, occasional events that seldom appear, even only once, but, in any case, it is necessary to know them. Otherwise, DBSCAN  would show these cases as noisy cases, which is not adequate for the goal of this work. In Fig. \ref{fig:dbscan12h} (a), the detected WWTP operation modes are shown in the space of the 3 main principal components (1, 2 and 3), while Fig. \ref{fig:dbscan12h} (b) shows the same operation modes from another perspective: on the space of the principal components 5, 6 and 7. Here, 15 operation modes were detected, while most of the samples are logically clustered in \textsl{Operation Mode 0}, the rest are exceptional days that are considered as \textsl{Operation Modes} in order to know about similar situations in the near future. Fig.\ref{fig:dbscan12hMes} (a) shows the same data partition, but the sample colours are in accordance with the month of the samples; while in the blow-up of the \textsl{Operation Mode 0}, it is possible to see some distribution patterns over months.

But all this has to be explained in terms of the WWTP domain, and therefore according to the variables managed by the WWTP managers. Tables \ref{tab:SHAP12ha} and \ref{tab:SHAP12h} show an extract containing the variables of the facility with the greatest impact for every operation mode detected by DBSCAN, and the units of the WWTP facility involved.

\begin{table}[b]
    \centering
    \makebox[\textwidth][c]{
     \resizebox{19cm}{!} {
    \begin{tabular}{|c|c|c|c|c|c|c|}    \hline
       \multicolumn{7}{|c|}{Operation modes} \\ \hline
        0 									& 1 						& 2 						& 3 								& 4 							& 5 							& 6 \\ \hline
        \cellcolor{PreThi}BP3042\_V.K\_Factor  				& \cellcolor{PreThi}PIDNE306.PID.KI  	& \cellcolor{PreThi}FD3044\_V.Q\_Max  & \cellcolor{PumArr}DGP5000\_EA\_VT.VPE  		& \cellcolor{PumArr}DGP5000\_EA\_VT.VPE  	& \cellcolor{Bio}IP223\_EA\_IA.VSca  		& \cellcolor{PreThi}FD3038\_V.Q\_Max  \\ \hline
        \cellcolor{PreThi}BP3041\_V.K\_Factor  				& \cellcolor{PreThi}FD3043\_V.Q\_Max  & \cellcolor{PreThi}FD3038\_V.Q\_Max  & \cellcolor{PumArr}DGP5000\_EA\_IT.Comunica  	& \cellcolor{PreThi}Impulsion.SP\_Paro  		& \cellcolor{PreThi}Impulsion.SP\_Paro  		& \cellcolor{PreThi}FD3043\_V.Q\_Max  \\ \hline
        \cellcolor{Bio}FE201.VSca  						& \cellcolor{PreThi}FD3036\_V.Q\_Max  & \cellcolor{PreThi}PIDNE306.PID.KI  	& \cellcolor{PumArr}DNT5026\_EA\_NT.Comunica  	& \cellcolor{PreThi}Impulsion.SP\_Arr  		& \cellcolor{PreThi}Impulsion.SP\_Arr 		& \cellcolor{PreThi}PIDNE306.PID.KD  \\ \hline
        \cellcolor{Unk}ZHRecirculacionFangosBiologico3.TotalActual  	& \cellcolor{PreThi}FD3044\_V.Q\_Max  & \cellcolor{PreThi}PIDNE306.PID.DB  	& \cellcolor{PumArr}DGP5000\_EA\_VT.Comunica 	 & \cellcolor{PreThi}Esp\_Gen.SP\_EV3032\_R & \cellcolor{PreThi}Esp\_Gen.SP\_EV3032\_R  & \cellcolor{PreThi}FD3035\_V.Q\_Max  \\ \hline
        \cellcolor{Dra}Escurridos.PT100.VALOR  				& \cellcolor{PreThi}FD3038\_V.Q\_Max  & \cellcolor{PreThi}FD3043\_V.Q\_Max  & \cellcolor{PumArr}GB.Comunica  			& \cellcolor{PreThi}Tormentas.Ton1001  	& \cellcolor{PreThi}PID1091.PID.KD  		& \cellcolor{PreThi}PIDNE306.PID.KI  \\ \hline
        \cellcolor{Bio}FR204\_EA\_IA.VSca 				& \cellcolor{PreThi}FD3035\_V.Q\_Max  & \cellcolor{PreThi}FD3036\_V.Q\_Max  & \cellcolor{PreThi}Impulsion.SP\_Paro  		& \cellcolor{PreThi}PID1090.PID.DB  		& \cellcolor{PreThi}PID\_BP3041.PID.MAXO  	& \cellcolor{PreThi}FD3037\_V.Q\_Max  \\ \hline
        \cellcolor{Bio}RE2081\_SA\_SAF.val  				& \cellcolor{PreThi}PIDNE306.PID.KD  	& \cellcolor{PreThi}FD3145\_V.Q\_Max  & \cellcolor{PreThi}Impulsion.SP\_Arr  			& \cellcolor{PreThi}Tormentas.SPL\_IP101  	& \cellcolor{PreThi}Tormentas.T1on1004  	& \cellcolor{PreThi}PIDNE306.PID.DB  \\ \hline
        \cellcolor{Bio}FR202\_EA\_IA.VSca  				& \cellcolor{PreThi}FD3037\_V.Q\_Max  & \cellcolor{PreThi}FD3035\_V.Q\_Max  & \cellcolor{PreThi}Esp\_Gen.SP\_EV3032\_R  	& \cellcolor{PreThi}Tormentas.SPL\_NP102  	& \cellcolor{PreThi}PID\_BP3042.PID.KI  	& \cellcolor{PreThi}FD3036\_V.Q\_Max  \\ \hline
        \cellcolor{Unk}ZHEntradaBiologico1.TotalActual  		& \cellcolor{PreThi}FD3145\_V.Q\_Max  & \cellcolor{PreThi}FD3037\_V.Q\_Max  & \cellcolor{PreThi}Tormentas.Ton1001  		& \cellcolor{PreThi}Tormentas.T1on1004  	& \cellcolor{PreThi}PID1091.PID.KI  		& \cellcolor{PreThi}FD3044\_V.Q\_Max  \\ \hline
        \cellcolor{Unk}ZHRecirculacionFangosBiologico1.TotalActual  	& \cellcolor{PreThi}PIDNE306.PID.DB  	& \cellcolor{PreThi}PIDNE306.PID.KD  	& \cellcolor{PreThi}Tormentas.SPL\_NP102  		& \cellcolor{PreThi}Tormentas.SPH\_NP102  	& \cellcolor{PreThi}Impulsion.SP\_Var\_Vaciado &\cellcolor{PreThi} FD3145\_V.Q\_Max  \\ \hline
    \end{tabular}
    }	}
    \caption{DBSCAN Operation Modes explained by SHAP ($Ts=12 hours$).}
		\label{tab:SHAP12ha}
\end{table}

\begin{table}[h]
    \centering
    \makebox[\textwidth][c]{
     \resizebox{19cm}{!} {
    \begin{tabular}{|c|c|c|c|c|c|c|c|}
    \hline
       \multicolumn{8}{|c|}{Operation modes} \\ \hline
        7 					     & 8 							& 9 							& 10 									& 11 							& 12 						& 13 									& 14 \\ \hline
        \cellcolor{PreThi}FD3037\_V.Q\_Max  & \cellcolor{Bio}IP223\_EA\_IA.VSca  		& \cellcolor{Bio}RIR1\_PID.PID.Q  		& \cellcolor{PreThi}EV3046V.on  					& \cellcolor{PumArr}DBA5006.TBVA.TL  		& \cellcolor{PreThi}Impulsion.SP\_Aut  	& \cellcolor{PumArr}GB.Comunica  				& \cellcolor{Bio}RE2082\_EA\_IA.VSca  \\ \hline
        \cellcolor{PreThi}PIDNE306.PID.KI      & \cellcolor{Unk}ZHCaudalEntrada.TotalActual  	& \cellcolor{Bio}RIR1\_PID.PID.DB  		& \cellcolor{PreThi}DN3049A.est  				& \cellcolor{PumArr}DBA5006.TRAAR.TL  	& \cellcolor{PreThi}PID1090.VP  		& \cellcolor{PumArr}DGP5000\_EA\_IT.Comunica  		& \cellcolor{PumArr}GB.VP  \\ \hline
        \cellcolor{PreThi}PIDNE306.PID.DB     & \cellcolor{PreThi}IP101.VSca  			& \cellcolor{Bio}RIR4\_PID.PID.DB  		& \cellcolor{PreThi}EV3147V.on  					& \cellcolor{PumArr}DBA5006.TBAR.TL  	& \cellcolor{Dig}IQ4061.VSca  		& \cellcolor{PumArr}DGP5000\_EA\_VT.Comunica  		& \cellcolor{PumArr}GB1.E\_VP  \\ \hline
        \cellcolor{PreThi}FD3043\_V.Q\_Max  & \cellcolor{PumArr}GB.E\_CF  			& \cellcolor{Bio}RIR2\_PID.PID.DB  		& \cellcolor{PreThi}IT3089.VSca  					& \cellcolor{PumArr}DBA5005.TRAVA.TL  	& \cellcolor{PreThi}FP107.VSca  		& \cellcolor{PumArr}DNT5026\_EA\_NT.Comunica  		& \cellcolor{PumArr}DNT5026.VSca  \\ \hline
        \cellcolor{PreThi}FD3038\_V.Q\_Max  & \cellcolor{Unk}AR110.VSca  			& \cellcolor{Bio}RI2054A\_SA\_SAF.val 		& \cellcolor{Dig}FC4002.Q\_Max  				& \cellcolor{PumArr}DBA5007.TBAR.TL  	& \cellcolor{Dig}IQ4065.VSca  		& \cellcolor{Unk}ControlPICs.Noche\_Toff4  			& \cellcolor{Bio}RE2082\_SA\_SAF.val  \\ \hline
        \cellcolor{PreThi}FD3044\_V.Q\_Max  & \cellcolor{PumArr}DNT5026.VSca  			& \cellcolor{Bio}RIR2\_PID.PID.KI  		& \cellcolor{Dig}L1S4008.est  					& \cellcolor{PumArr}DBA5007.TCMKAR.TL  	& \cellcolor{Bio}VR2068\_EA\_IA.VSca  & \cellcolor{Unk}PRE.MINUTo\_ENCENDIDO1 	& \cellcolor{PumArr}GB1.NBO  \\ \hline
        \cellcolor{PreThi}FD3145\_V.Q\_Max  & \cellcolor{PumArr}GB1.E\_VP  			& \cellcolor{Bio}RIR4\_PID.PID.MAXO  		& \cellcolor{Dig}FC4001.Q\_Max  				& \cellcolor{PumArr}DBA5007.TCMAR.TL  	& \cellcolor{Bio}FE201.VSca  		& \cellcolor{Unk}PRE.MINUTO\_APAGADO1  	& \cellcolor{Bio}IP223\_EA\_IA.VSca  \\ \hline
        \cellcolor{PreThi}FD3035\_V.Q\_Max  & \cellcolor{PumArr}GB.VP  				& \cellcolor{Bio}RIR3\_PID.PID.Q  		& \cellcolor{Dig}FC4004.Q\_Max  				& \cellcolor{PumArr}DBA5005.TCMVA.TL  	& \cellcolor{Bio}VR2066\_EA\_IA.VSca  & \cellcolor{Unk}ControlPICs.Lluvia\_Ton1  			& \cellcolor{Unk}ZHCaudalEntrada.TotalActual  \\ \hline
        \cellcolor{PreThi}PIDNE306.PID.KD     & \cellcolor{PumArr}DGP5000\_EA\_IT.VPE  	& \cellcolor{Bio}RIR3\_PID.PID.KI  		& \cellcolor{Dig}FC4003.Q\_Max  				& \cellcolor{PumArr}DBA5005.TRAAR.TL  	& \cellcolor{Bio}VR2071\_EA\_IA.VSca  & \cellcolor{Unk}ControlPICs.Lluvia\_Ton3  			& \cellcolor{Unk}AR109.VSca  \\ \hline
        \cellcolor{PreThi}FD3036\_V.Q\_Max  & \cellcolor{Unk}AR109.VSca  			& \cellcolor{Dra}Escurridos.QIT102.VALOR  	& \cellcolor{Unk}DIG.MINUTO\_ENCENDIDO1  	& \cellcolor{PumArr}DBA5006.TRAVA.TL  	& \cellcolor{Bio}IP223\_EA\_IA.VSca  	& \cellcolor{Unk}ControlPICs.Lluvia\_Ton2  			& \cellcolor{Unk}AR110.VSca  \\ \hline
    \end{tabular}
    }}
    \caption{DBSCAN Operation Modes explained by SHAP ($Ts=12 hours$).}
		\label{tab:SHAP12h}
\end{table}

In this way, all the information from the monitoring systems of the WWTP, thousands of variables, is summarized as operation modes of the facility. Each one can be located at a glance as a point in a low-dimensional variable space space, manageable by technicians in charge of the plant. Their closeness regarding other known WWTP operation modes and the impact of every monitored variable and corresponding WWTP unit can also be easily seen.

These results were also checked with the records of the WWTP facility in order to explain the unsupervised modes obtained by the available expert knowledge: \textsl{Operation Mode 0}, the  most populated, concerns the regular operation of the plant over the year; while \textsl{Operation Modes 4 and 13} match with shutdown standbys of the plant due to maintenance and issues on power lines according to expert knowledge and records. Moreover, an arrival pumping  variable is detected by SHAP for both as an explanation for these exceptional operation modes. \textsl{Operation Modes 5, 8, 9 and 14} concern episodes of heavy rain at the end of summer, in autumn, winter and summer respectively. Therefore, variables related to the rain are highlighted by SHAP. Finally, \textsl{Operation Modes 1, 2, 6, 7 and 12} seem to be some type of sporadic disturbances at the edge of the regular operation. Here SHAP highlights the preprocessing and thickening facilities variables; while \textsl{Operation Modes 3, 10 and 11} are more distanced from the regular operation mode, while modes $3$ and $11$ share the influence of the pumping arrival facility variables according to SHAP.

\vspace{0.5\baselineskip}
\textbf{ $\bullet$ Case: $T_s=12 hours$}, Online
\vspace{0.5\baselineskip}

Here, every new sample is managed in this mode model, checking the relationship regarding the known WWTP operation modes; while their nature regarding the WWTP variables is explained by SHAP. In short, each new sampling is turned into a point in the low-dimensional mode space and linked with operation modes known so far, being explained by the WWTP variables that have the greatest influence for this location.

\begin{figure}[t]
	\centering
    	\makebox[\textwidth][c]{
	\begin{tabular}{cc}
			\includegraphics[height=8cm]{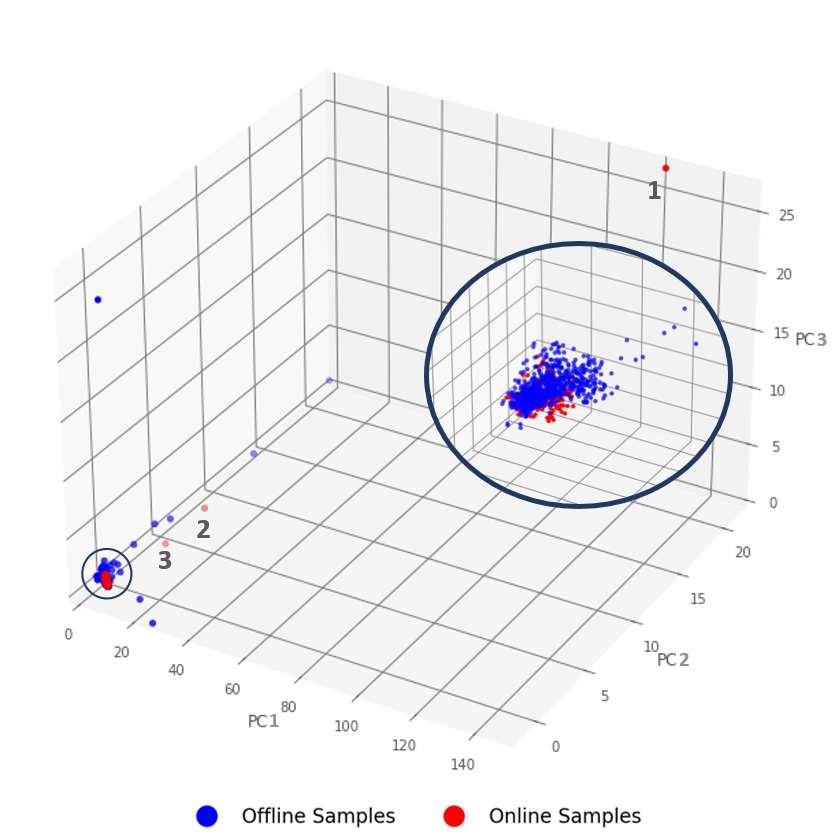} & \includegraphics[height=8cm]{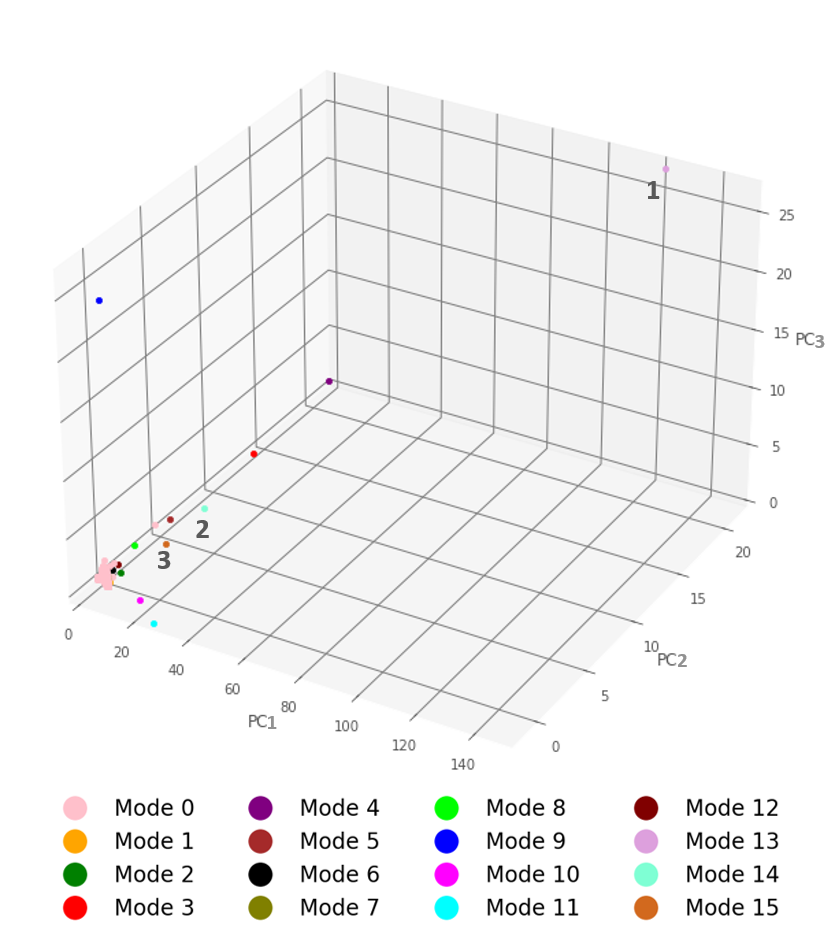}\\
			(a) Online samples into WWTP Operation Mode space.					& (b) IncrementalDBSCAN Operation Modes.
	\end{tabular}
	}
	\caption{Online WWTP Operation Model ($T_s=12 hours$) }
	\label{fig:dbscan12hMesOn}

\end{figure}

Three months of data samples were used for this stage, being processed as in the previous step. Fig. \ref{fig:dbscan12hMesOn} (a) shows the online samples in red for the PC1, PC2 and PC3 input space. As can be observed in both Fig. \ref{fig:dbscan12hMesOn} (a) and (b), most of these new samples are in the space of the \textsl{Operation Mode 0}, the standard operation mode of the WWTP plant; while others, which are not linked to the known operation modes, such as the samples marked as $1$, $2$ and $3$, are not known regarding the historical data. In Table \ref{tab:samplesonvar}, these new modes are explained by the WWTP variables  and units involved. Samples number $1$ and $2$ match with a general shutdown for 2 hours due to the change of power lines; so the most influential variables according to SHAP are the same, corresponding to the digestion and biological facilities. On the other hand, sample number $3$ concerns a connection failure, and therefore the apperance of a new variable related to the lighting.

\begin{table}[H]
    \centering
     \resizebox{13cm}{!} {
    \begin{tabular}{|c|c|c|}
       \hline
       \multicolumn{3}{|c|}{WWTP Relevant Variables for New Unknown online modes } \\ \hline
        1							& 2 						& 3  \\  \hline \hline
        \cellcolor{Dig} FC4003.Q\_Max  		& \cellcolor{Dig}L1S4008.est  		& \cellcolor{Dig}L1S4008.est  \\ \hline
        \cellcolor{Dig} L1S4008.est  		& \cellcolor{Dig}FC4002.Q\_Max  	& \cellcolor{Dig}FC4001.Q\_Max  \\ \hline
        \cellcolor{Dig}FC4002.Q\_Max  		& \cellcolor{Dig}FC4001.Q\_Max  	& \cellcolor{Dig}FC4003.Q\_Max  \\ \hline
        \cellcolor{Dig}FC4001.Q\_Max  		& \cellcolor{Dig}FC4004.Q\_Max  	& \cellcolor{Dig}FC4004.Q\_Max  \\ \hline
        \cellcolor{Dig}FC4004.Q\_Max  		& \cellcolor{Dig}FC4003.Q\_Max  	& \cellcolor{Dig}FC4002.Q\_Max  \\ \hline
        \cellcolor{Bio}RIR4\_PID.PID.Q  		& \cellcolor{Bio}RIR1\_PID.PID.Q  	& \cellcolor{Unk}DIG.MINUTO\_ENCENDIDO1  \\ \hline
        \cellcolor{Bio}RIR2\_PID.PID.MAXO  	& \cellcolor{Bio}RIR1\_PID.PID.KI  	& \cellcolor{Bio}RIR3\_PID.PID.MAXO  \\ \hline
        \cellcolor{Bio}RIR3\_PID.PID.KI  		& \cellcolor{Bio}RIR2\_PID.PID.MAXO  	& \cellcolor{Bio}RIR1\_PID.PID.KI  \\ \hline
        \cellcolor{Bio}RIR4\_PID.PID.KI  		& \cellcolor{Bio}RI2063A\_SA\_SAF.val  &\cellcolor{Bio} RI2054\_5\_6.T\_ON  \\ \hline
        \cellcolor{Bio}FA2128.cm  			& \cellcolor{Bio}RI2060A\_SA\_SAF.val  &\cellcolor{Bio} RIR3\_PID.PID.Q  \\ \hline

    \end{tabular}
     }

    \caption{WWTP variable relevance for new online operation modes ($T_s=12 hours$).}
	\label{tab:samplesonvar}
\end{table}

\begin{figure}
   	 \centering
	   \makebox[\textwidth][c]{
			\includegraphics[height=8.5cm]{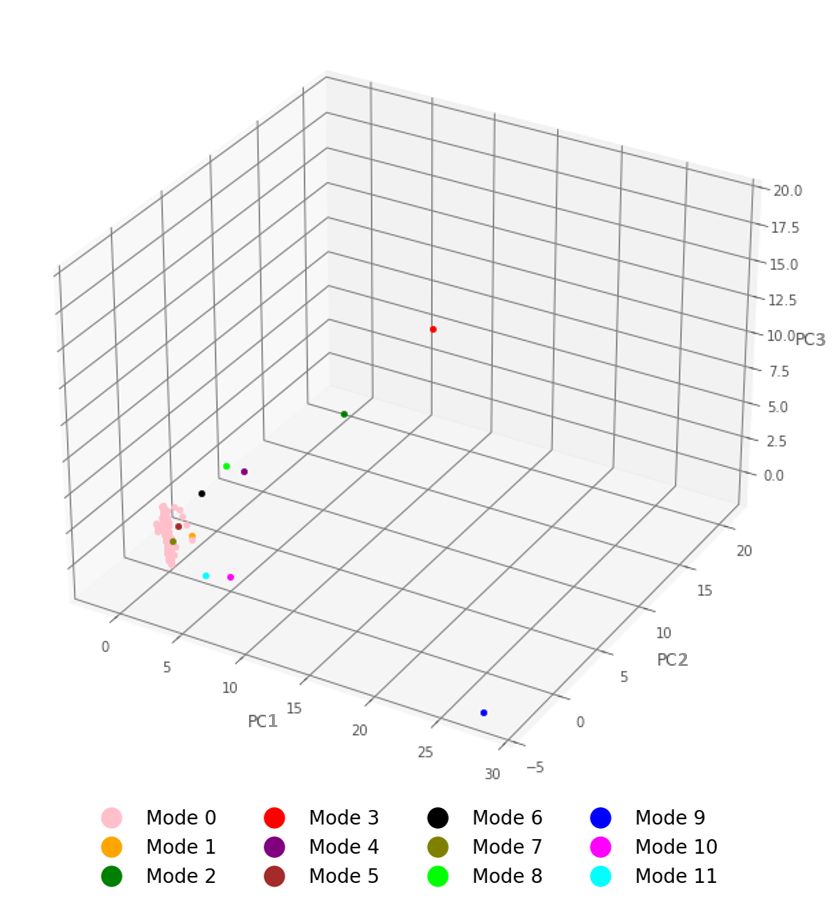}
	}
	\caption{Updated WWTP Operation Mode model.}
	\label{fig:dbscan12hMesOnUp}
\end{figure}

When the WWTP operation mode model  is considered outdated, then it can be  rebuilt  following the offline steps, generating a new WWTP operation model as shown in Fig. \ref{fig:dbscan12hMesOnUp}, taking into account 15 months of historical samples from the WWTP. Here, some previously exceptional mode samples have now transitioned into \textsl{Operation Mode 0} in this new scenario. Conversely, the majority of new samples fall into \textsl{Operation Mode 0}, while $3$ new exceptional operation modes emerge. Once again, these results have been validated using available expert knowledge, supporting the obtained unsupervised operation modes.

\vspace{0.5\baselineskip}
\textbf{ $\bullet$ Case: $T_s=24 hours$}, Offline
\vspace{0.5\baselineskip}

Here, a similar analysis has been made for the sample time $T_s=24 hours$. Once more, the parameter $MinPoints$ was selected as only $1$ point in order to allow a better detection of minimal  clusters, due to the WWTP operation. Then,  $\varepsilon$ is tuned at $1.1$, following the tuning methodology above described. 

\begin{figure}[h]
   	 \centering
	   \makebox[\textwidth][c]{
	\begin{tabular}{cc}
			\includegraphics[height=8.5cm]{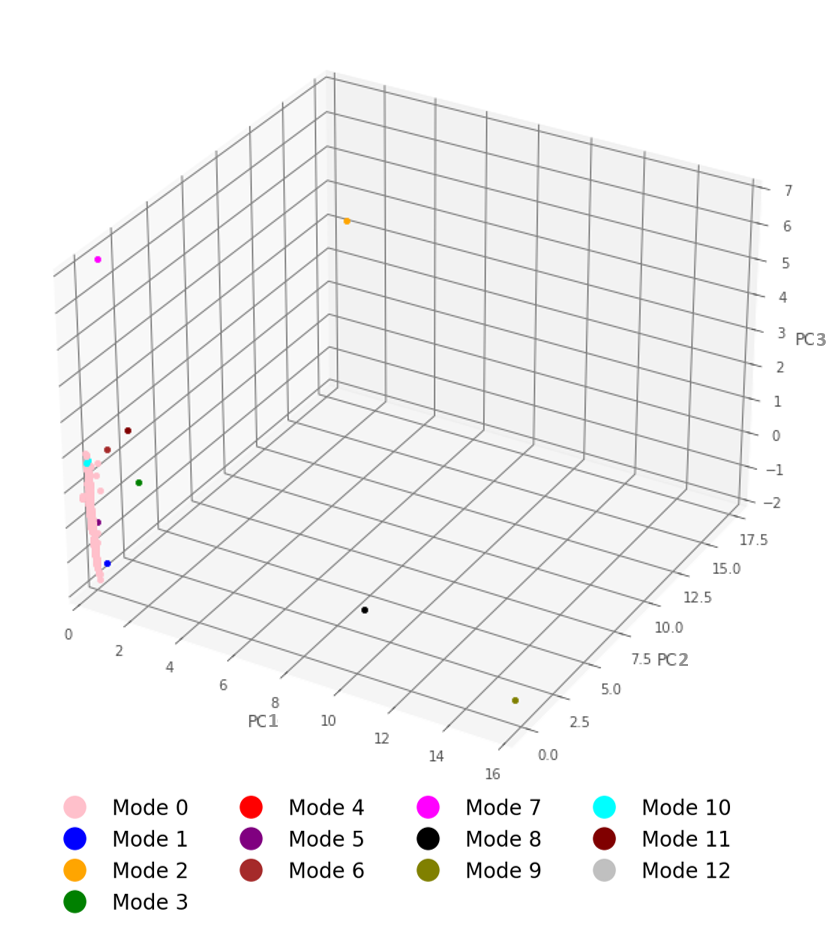} & \includegraphics[height=8.5cm]{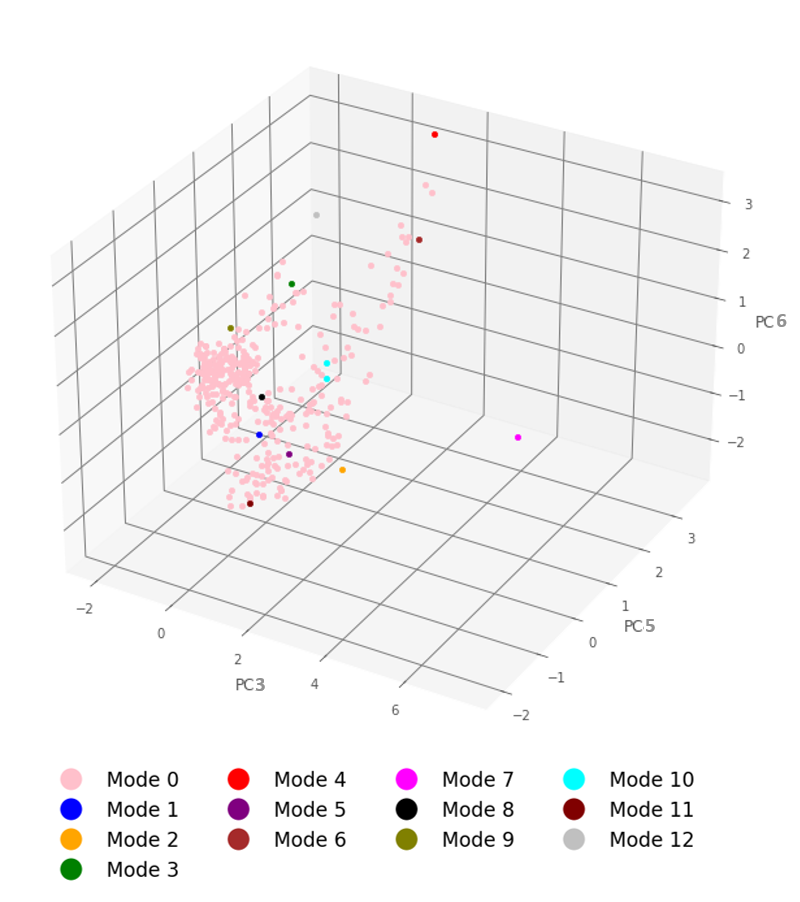}\\
			(a) Operation Modes: view by  $PC1$,$PC2$ and $PC3$.					& (b) Operation Modes: view by $PC3$, $PC5$ and $PC6$.
	\end{tabular}
	}
	\caption{Operation Modes based on DBSCAN ($\varepsilon=1.1$  and the $MinPoints=1$).}
	\label{fig:dbscan24h}
\end{figure}

In Fig. \ref{fig:dbscan24h} (a), the detected WWTP operation modes are presented in the space of the three main principal components (extracted variables); while Fig. \ref{fig:dbscan24h} (b) illustrates these same  operation modes in the space of principal components 3, 5, and 6, to improve the visibility of the operation modes regarding \textsl{Operation Mode 0}. 

Now, 13 operation modes have been detected: \textsl{Operation Mode 0}, into which almost all the samples are clustered, and the rest of the modes, which correspond to exceptional days according to the experts. 

\begin{figure}[H]
   	 \centering
	   \makebox[\textwidth][c]{
	\begin{tabular}{cc}
			\includegraphics[height=8.5cm]{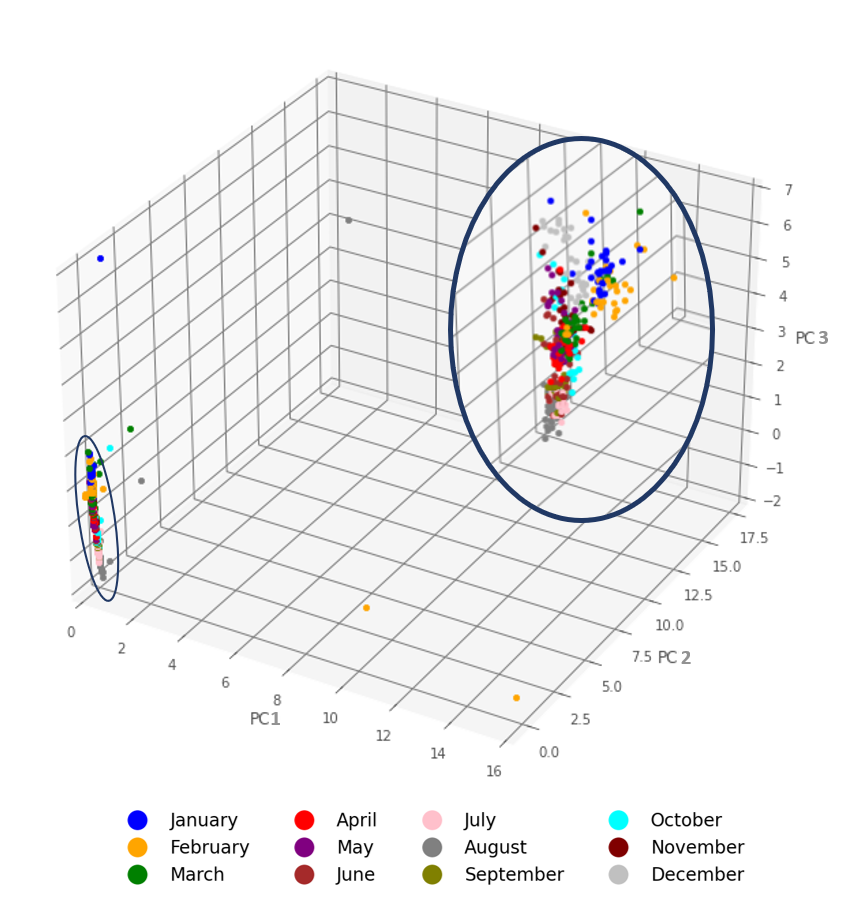} & \includegraphics[height=8.5cm]{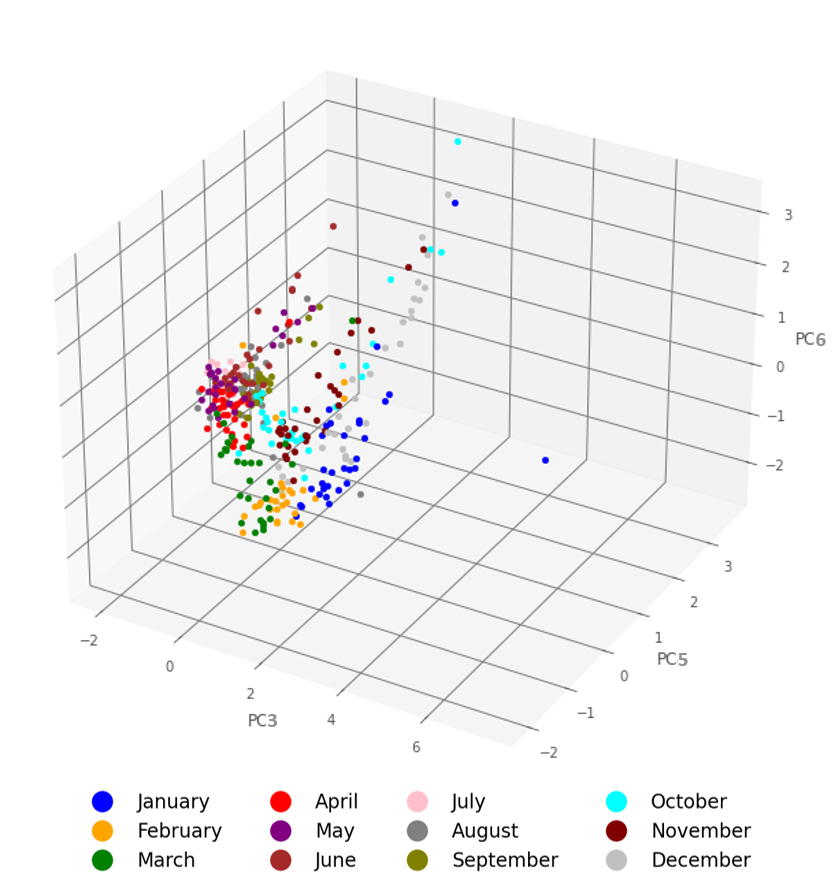}\\
			(a) Operation Modes by months ($PC1$,$PC2$ and $PC3$).					& (b) Operation  Modes by months ($PC3$, $PC5$ and $PC6$).
	\end{tabular}
}
	\caption{Operation Modes based on months ($\varepsilon=1.1$  and the $MinPoints=1$)}
	\label{fig:dbscan24hMes}

\end{figure}

Under the WWTP operations, these points are not considered noise, as they show different situations in which the plant was working in comparison to its standard operation mode over the year. Fig. \ref{fig:dbscan24hMes} (a) shows the same as in Fig. \ref{fig:dbscan24h} (a), but here every sample is in colour according to its sample month, in order to visualize some patterns based on this time feature. Moreover, Fig. \ref{fig:dbscan24hMes} (b) shows the modes in the 3, 5 and 6 principal component space for a better discrimination of the \textsl{Mode 0}.

The explanation of these modes is newly obtained by SHAP, and shown in Tables \ref{tab:SHAP24ha} and \ref{tab:SHAP24h}. The cell colour is associated with the WWTP unit for each variable. All this allows us to know the operation modes as well as the facility variables and units involved, which allows us to locate a WWTP data sample in the operation mode model and its match regarding the known modes over time at any given moment.

\begin{table}[h]
    \centering
    \makebox[\textwidth][c]{
     \resizebox{19cm}{!} {
    \begin{tabular}{|c|c|c|c|c|c|c|}    \hline
       \multicolumn{7}{|c|}{Operation modes} \\ \hline
        0 							& 1 						& 2 									& 3 									& 4 							& 5 						& 6 \\ \hline
        \cellcolor{Bio}RE2081\_SA\_SAF.val  		& \cellcolor{PreThi}FD3036\_V.Q\_Max  & \cellcolor{Unk}ControlPICs.Noche\_Toff4  			& \cellcolor{Unk}ZHEntradaBiologico1.TotalActual  		& \cellcolor{PumArr}DNT5026.VSca  		& \cellcolor{PreThi}PIDNE306.PID.KI  	& \cellcolor{Bio}IP223\_EA\_IA.VSca  \\ \hline
        \cellcolor{PreThi}BP3042\_V.K\_Factor  		& \cellcolor{PreThi}FD3037\_V.Q\_Max  & \cellcolor{PreThi}Impulsion.SP\_Paro  			&\cellcolor{PumArr} DBA5006.TEMP\_COJINETE\_SOPORTE & \cellcolor{PumArr}GB1.E\_VP  			& \cellcolor{PreThi}FD3038\_V.Q\_Max  & \cellcolor{Unk}ZHCaudalEntrada.TotalActual  \\ \hline
        \cellcolor{Bio}FE201.VSca  				& \cellcolor{PreThi}FD3043\_V.Q\_Max  & \cellcolor{PreThi}Impulsion.SP\_Arr  				& \cellcolor{Unk}AR111.VSca  					& \cellcolor{PumArr}GB.VP  			& \cellcolor{PreThi}PIDNE306.PID.DB  	& \cellcolor{PreThi}IP101.VSca  \\ \hline
        \cellcolor{Dig}IQ4103.VSca  			& \cellcolor{PreThi}FD3145\_V.Q\_Max  & \cellcolor{Unk}PRE.MINUTO\_APAGADO1  	& \cellcolor{PumArr}DBA5006.TEMP\_STATOR\_PH\_1 	& \cellcolor{Unk}ZHCaudalEntrada.TotalActual & \cellcolor{PreThi}PIDNE306.PID.KD  	& \cellcolor{PumArr}GB.E\_CF  \\ \hline
        \cellcolor{PreThi}BP3041\_V.K\_Factor  		& \cellcolor{PreThi}FD3035\_V.Q\_Max  & \cellcolor{Unk}PRE.MINUTO\_ENCENDIDO1  	& \cellcolor{PumArr}DBA5006.TEMP\_STATOR\_PH\_2  	& \cellcolor{PreThi}IP101.VSca  			& \cellcolor{PreThi}FD3044\_V.Q\_Max  & \cellcolor{Unk}AR110.VSca  \\ \hline
        \cellcolor{Unk}ZHFangosExceso.TotalActual  	& \cellcolor{PreThi}PIDNE306.PID.KI  	& \cellcolor{Unk}ControlPICs.Lluvia\_Ton1  			& \cellcolor{PumArr}DBA5006.TEMP\_STATOR\_PH\_3  	& \cellcolor{PumArr}GB.E\_CF  			& \cellcolor{PreThi}FD3035\_V.Q\_Max  & \cellcolor{PumArr}GB1.E\_VP  \\ \hline
        \cellcolor{Bio}RIR4\_PID.VC  			& \cellcolor{PreThi}PIDNE306.PID.DB  	& \cellcolor{PreThi}DecanPrim.T\_R1090  			& \cellcolor{PumArr}DGP5000\_EA\_IT.VPE  			& \cellcolor{PumArr}DGP5000\_EA\_IT.VPE  	& \cellcolor{PreThi}FD3145\_V.Q\_Max  & \cellcolor{PumArr}DNT5026.VSca  \\ \hline
        \cellcolor{Dig}IQ4064.VSca 		 		& \cellcolor{PreThi}PIDNE306.PID.KD  	& \cellcolor{Unk}ControlPICs.Lluvia\_Ton3  			& \cellcolor{Unk}AM4101.VSca  					& \cellcolor{Bio}IP223\_EA\_IA.VSca  		& \cellcolor{PreThi}FD3043\_V.Q\_Max  & \cellcolor{PumArr}GB.VP  \\ \hline
        \cellcolor{Unk}AM4103.VSca  			& \cellcolor{PreThi}FD3038\_V.Q\_Max  & \cellcolor{Unk}ControlPICs.Lluvia\_Toff4 			& \cellcolor{Bio}IP223\_EA\_IA.VSca  				& \cellcolor{Unk}AR110.VSca  			& \cellcolor{PreThi}FD3037\_V.Q\_Max  & \cellcolor{PumArr}DGP5000\_EA\_IT.VPE  \\ \hline
        \cellcolor{Bio}RE2080\_SA\_SAF.val  		& \cellcolor{PreThi}FD3044\_V.Q\_Max  & \cellcolor{Unk}ControlPICs.Lluvia\_Ton2  			& \cellcolor{Unk}ZHFangosExceso.TotalActual  		& \cellcolor{Unk}AR109.VSca  			& \cellcolor{PreThi}FD3036\_V.Q\_Max  & \cellcolor{Unk}AR109.VSca  \\ \hline
    \end{tabular}
    }	}	
    \caption{DBSCAN Operation Modes explained by SHAP ($Ts=24 hours$).}
		\label{tab:SHAP24ha}
\end{table}

\begin{table}[h]
    \centering
    \makebox[\textwidth][c]{
     \resizebox{19cm}{!} {
    \begin{tabular}{|c|c|c|c|c|c|}
    \hline
       \multicolumn{6}{|c|}{Operation modes} \\ \hline
        7 								& 8 							& 9 							& 10 							& 11 									& 12 \\ \hline
        \cellcolor{Bio}RI2063A\_SA\_SAF.val  			& \cellcolor{PreThi}EV3147V.on  			& \cellcolor{PumArr}DBA5005.TCMVA.TL  	& \cellcolor{PumArr}GB.VP  			& \cellcolor{PumArr}DGP5000\_EA\_VT.Comunica  		& \cellcolor{Bio}RE2082\_EA\_IA.VSca  \\ \hline
        \cellcolor{Bio}RIR3\_PID.PID.KI  				& \cellcolor{PreThi}EV3046V.on  			& \cellcolor{PumArr}DBA5006.TRAAR.TL  	& \cellcolor{PumArr}GB1.E\_VP 			& \cellcolor{PumArr}DGP5000\_EA\_IT.Comunica  		& \cellcolor{Bio}RE2082\_SA\_SAF.val  \\ \hline
        \cellcolor{Bio}RIR2\_PID.PID.DB  				& \cellcolor{PreThi}DN3049A.est  		& \cellcolor{PumArr}DBA5006.TCMKAR.TL  	& \cellcolor{PumArr}DNT5026.VSca  		& \cellcolor{PumArr}GB.Comunica  				& \cellcolor{PumArr}GB.VP  \\ \hline
        \cellcolor{Bio}RIR1\_PID.PID.KI  				& \cellcolor{PreThi}IT3089.VSca  			& \cellcolor{PumArr}DBA5006.TRAVA.TL  	& \cellcolor{PumArr}DGP5000\_EA\_IT.VPE  	& \cellcolor{PumArr}DNT5026\_EA\_NT.Comunica  		& \cellcolor{PumArr}DNT5026.VSca  \\ \hline
        \cellcolor{Unk}BIO.MINUTO\_ENCENDIDO1  & \cellcolor{PumArr}DBA5002.TRAAR.TL  	& \cellcolor{PumArr}DBA5006.TCMKVA.TL  	& \cellcolor{PumArr}GB1.NBO  			& \cellcolor{Unk}ControlPICs.Noche\_Toff4  			& \cellcolor{PumArr}GB1.E\_VP  \\ \hline
        \cellcolor{Dra}Escurridos.QIT102.VALOR  		& \cellcolor{PumArr}DBA5011.TRAAR.TL 	& \cellcolor{PumArr}DBA5006.TBVA.TL  		& \cellcolor{PreThi}Impulsion.SP\_Aut  		& \cellcolor{Unk}PRE.MINUTO\_ENCENDIDO1  	& \cellcolor{Bio}IP223\_EA\_IA.VSca  \\ \hline
        \cellcolor{Bio}RIR3\_PID.PID.DB  				& \cellcolor{PumArr}DNT5026\_EA\_NT.K\_H  	& \cellcolor{PumArr}DBA5005.TCMKAR.TL  	& \cellcolor{Dig}IQ4065.VSca  			& \cellcolor{Unk}PRE.MINUTO\_APAGADO1  	& \cellcolor{PumArr}GB1.NBO  \\ \hline
        \cellcolor{Bio}RIR4\_PID.PID.MAXO  			& \cellcolor{PumArr}GB.SP\_CLH  		& \cellcolor{PumArr}DBA5007.TBAR.TL  	& \cellcolor{Bio}VR2071\_EA\_IA.VSca  		& \cellcolor{Unk}ControlPICs.Lluvia\_Ton3  			& \cellcolor{PumArr}DGP5000\_EA\_IT.VPE  \\ \hline
        \cellcolor{Bio}RIR1\_PID.PID.MAXO  			& \cellcolor{PumArr}DBA5010.TCMAR.TL  	& \cellcolor{PumArr}DBA5005.TRAAR.TL  	& \cellcolor{PreThi}FD307.VSca  			& \cellcolor{PreThi}DecanPrim.T\_R1090  			& \cellcolor{Unk}AR110.VSca  \\ \hline
        \cellcolor{Bio}RIR1\_PID.PID.Q  				& \cellcolor{PumArr}DBA5004.TRAVA.TL  	& \cellcolor{PumArr}DBA5005.TBAR.TL  	& \cellcolor{PreThi}FD308.VSca  			& \cellcolor{Unk}ControlPICs.Lluvia\_Ton1  			& \cellcolor{Unk}AR109.VSca  \\ \hline
    \end{tabular}
    }}
    \caption{DBSCAN Operation Modes explained by SHAP ($Ts=24 hours$).}
		\label{tab:SHAP24h}
\end{table}

These results are also checked regarding  the available expert knowledge: \textsl{Operation Mode 0} corresponds to the regular operation of the plant over the year, while \textsl{Operation Modes 2 and 11} concern a general plant shutdown for maintenance. On the other hand, \textsl{Operation Modes 3, 4, 6, 7, 10 and 12} are heavy rainy days, with $4$ and $6$ being the heaviest; consequently, the intake flow variables are  highly important for SHAP. Finally, \textsl{Operation Modes 1, 5, 8 and 9} seem to be an unknown event a short distance from the regular operation mode; while modes $1$ and $5$ seem to be a similar event because of the most important variables, related to the preprocessing and thickening facilities, as well as modes $8$ and $9$ which are related to the pumping arrival facility.

\vspace{0.5\baselineskip}
\textbf{ $\bullet$ Case: $T_s=24 hours$}, Online
\vspace{0.5\baselineskip}

Now, each new data sample is shown in this modes model and input space to know its relationship or similarity to known operation modes.

\begin{figure}[h]
   	 \centering
	   \makebox[\textwidth][c]{
	\begin{tabular}{cc}
			\includegraphics[height=8.5cm]{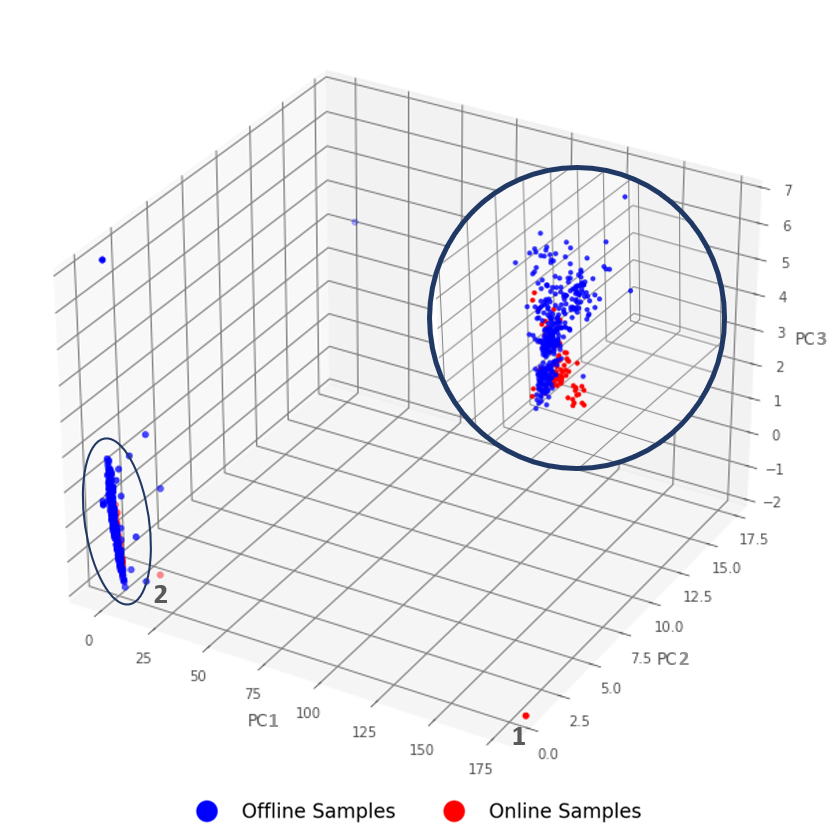}  & \includegraphics[height=8.5cm]{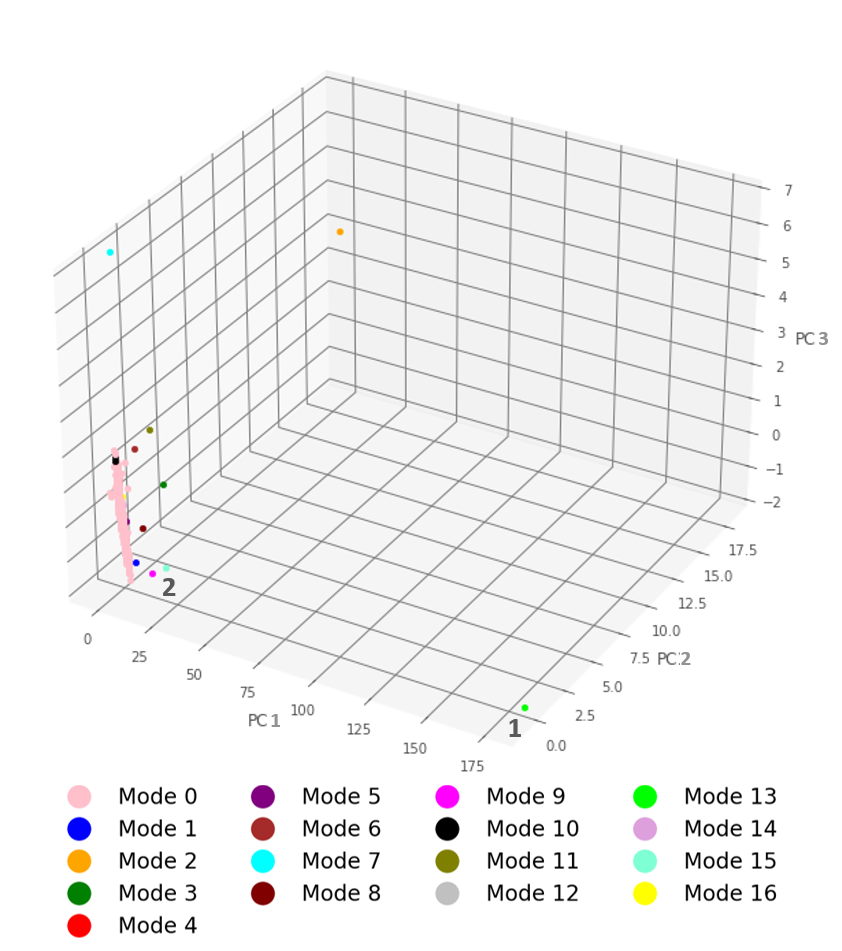}\\
			(a) Online samples into WWTP Operation Mode space.					& (b) IncrementalDBSCAN Operation Modes.
	\end{tabular}
	}
	\caption{Online WWTP Operation Model ($T_s=24  hours$)}
	\label{fig:dbscan24hMesOn}
\end{figure}

Three months of samples are used for this stage, processed in the same manner as in the previous step. Fig. \ref{fig:dbscan24hMesOn} (a) shows the offline modes in blue and online samples in red. The great majority of online samples go to \textsl{Operation Mode 0} (see Fig. \ref{fig:dbscan24hMesOn} (a) and (b)), while samples marked as $1$ and $2$ are far from the regular operation mode,  being explained by SHAP in Table \ref{tab:samplesonvar24h}. Sample number $1$ concerns a general plant shutdown which lasted 2 hours due to electrical matters, according to the WWTP experts. Meanwhile, its most influential variables according to SHAP are concerned with the biological units. On the other hand, sample number $2$ matches with a connection failure in the plant, with the pumping arrival facility being affected, according to SHAP.

\begin{table}[H]
    \centering
     \resizebox{9cm}{!} {
    \begin{tabular}{|c|c|}
       \hline
       \multicolumn{2}{|c|}{WWTP Variable Relevance for New Unknown on-line modes} \\ \hline
        1							& 2 \\  \hline \hline
        \cellcolor{Bio}RIR3\_PID.PID.DB  		& \cellcolor{PumArr}DBA5008.TTF\_H.AC  \\ \hline
        \cellcolor{Bio}RIR2\_PID.PID.MAXO  	& \cellcolor{PumArr}DBA5010.TBAR.TL  \\ \hline
        \cellcolor{Bio}RI2060A\_SA\_SAF.val  	& \cellcolor{PumArr}DBA5002.TBVA.TL  \\ \hline
        \cellcolor{Bio}RIR2\_PID.PID.DB  		& \cellcolor{PumArr}GB.TCPB.TL  \\ \hline
        \cellcolor{Bio}RIR4\_PID.PID.MAXO  	& \cellcolor{PumArr}DBA5003.TBVA.TL  \\ \hline
        \cellcolor{Bio}RIR3\_PID.PID.MAXO  	& \cellcolor{PumArr}GB.TPB.TL  \\ \hline
        \cellcolor{Bio}RIR1\_PID.PID.KI  		& \cellcolor{PumArr}DBA5008.TCMKVA.TL  \\ \hline
        \cellcolor{Bio}RIR4\_PID.PID.Q  		& \cellcolor{PumArr}DBA5003.TRAAR.TL  \\ \hline
        \cellcolor{Bio}RIR2\_PID.PID.KI  		& \cellcolor{PumArr}DBA5009.TBAR.TL  \\ \hline
        \cellcolor{Bio}RI2063A\_SA\_SAF.val  	& \cellcolor{PumArr}DBA5007.TRAVA.TL  \\ \hline

    \end{tabular}
	}
    \caption{WWTP  variable relevance  for new online operation modes. ($T_s=24 hours$)}
     \label{tab:samplesonvar24h}
\end{table}

\begin{figure}
   	 \centering
	   \makebox[\textwidth][c]{
			\includegraphics[height=8.5cm]{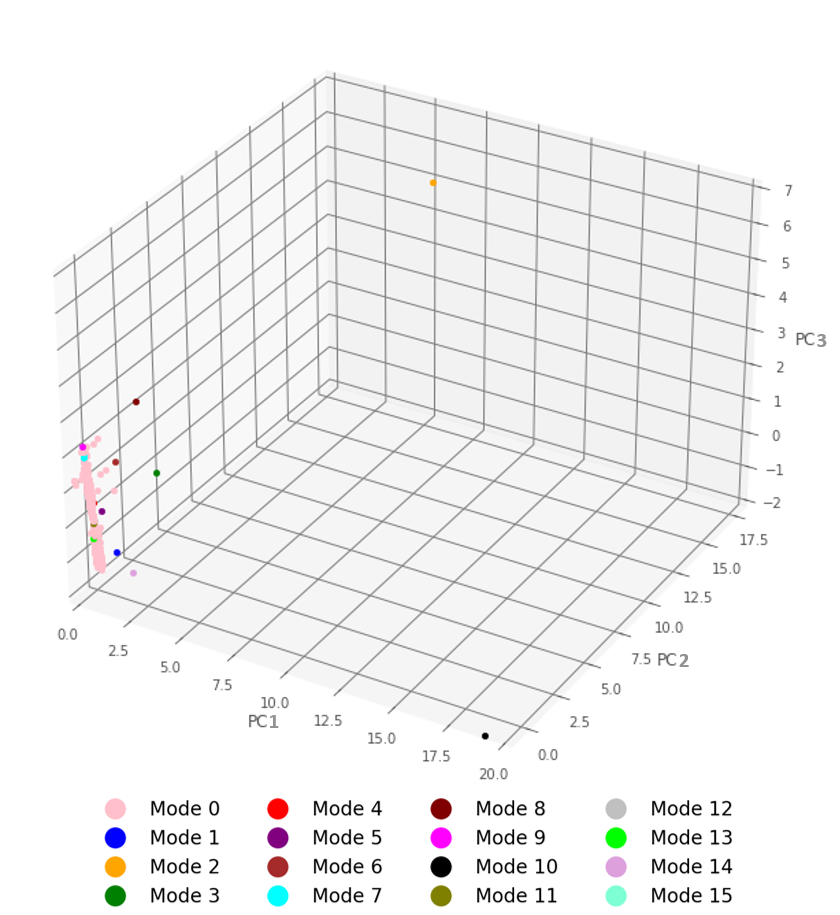} 
	}
	\caption{Updated WWTP Operation Mode model.}
	\label{fig:dbscan24hMesOnUp}

\end{figure}

If the WWTP operation modes model is in need of an update, this is carried out by the  offline steps of this proposal. Fig. \ref{fig:dbscan24hMesOnUp} shows a new WWTP operation model considering the 15 months of historical WWTP data samples. Regarding the DBSCAN operation modes obtained over 12 months, some exceptional mode samples of those months have become \textsl{Operation Mode 0} in this new scenario. On the other hand, most of the new samples belong to \textsl{Operation Mode 0} and five new exceptional operation modes have emerged.
Once more, these  results have been checked with the available expert knowledge, justifying the obtained unsupervised modes.

\section{Conclusions}
\label{sec:conclusions}

This work introduces an engineering approach for knowing  the WWTP operation modes applied to a real case, AQUAVALL WWTP in Valladolid (Spain). The goal is to translate the large and high-dimensional databases  from WWTP monitoring systems, such as  SCADA systems, into a point of a low-dimensional WWTP operation modes space, in which the WWTP operation modes are recorded over time as clusters, explained by their most influential WWTP variables.

All this would allow plant managers to know, at a glance, the standard operation modes, the new ones and  their closeness and similarities, as well as  the variables and units of the plant involved for each one. The large and  high-dimensional database is summarized in a small number of operation modes or points in a reduced input space. All this has been  carried out using the historical records of the monitoring system of the AQUAVALL WWTP over 15 months: 658084 records with  3231 monitored variables for each one.

Resampling, expert knowledge, PCA based dimensionality reduction, density based clustering DBSCAN, and SHAP values for explanation, are the compounds of the methodology applied to deal with the issues of the real WWTP case used in this work.

This methodology was tested with resampling times of 12 hours and 24 hours. It was possible to detect operation modes by matching them with the expert knowledge and the events saved in the plant records, such as shutdown, electrical issues and heavy rain. Explaining all these operation modes by the WWTP variables and units involved, and by analyzing SHAP explanations, we confirmed that the most influential process variables were consistent with expert knowledge, reinforcing the fact that the detected clusters were not arbitrary and reflected real operation modes.

On the other hand, the results revealed some slight differences between the two resampling times: the 12 hour sample time captured short-term variations in WWTP operations. The detailed patterns obtained are valuable for detecting transient states, sudden process fluctuations, and short-term anomalies. In contrast, while the 24 hour sample time can help in identifying stable operation modes, it may have obscured rapid changes that could be crucial for early anomaly detection and process optimization.

Moreover, the offline stage has allowed a comprehensive cluster analysis, while the online stage has demonstrated the applicability of the method in real-time monitoring. Furthermore, the necessity for model updating has been demonstrated because of the dynamic operation of the WWTP, where treatment processes can fluctuate due to changes in the influent characteristics, seasonal variations, and operation adjustments, changing the operation modes obtained with DBSCAN.

The work in progress is focused on the implementation of the methodology in the real WWTP to automate the operation mode detection and its interpretability, thus providing actionable insights for the operators.

\bibliographystyle{elsarticle-num} 
\bibliography{./Aquaexplain}

\end{document}